# Dynamic directed functional connectivity as a neural biomarker for objective motor skill assessment

Anil Kamat[1], Rahul Rahul[1], Anirban Dutta[6], Lora Cavuoto[2], Uwe Kruger[1], Harry Burke[4], Matthew Hackett[5], Jack Norfleet[5], Steven Schwaitzberg[3], Suvranu De[7]

[1]Center for Modeling, Simulation, and Imaging for Medicine, Rensselaer Polytechnic Institute, Troy, New York 12180, USA

[2]Department of Industrial and Systems Engineering, University at Buffalo, Buffalo, NY

[3]Department of Surgery, University at Buffalo Jacobs School of Medicine and Biomedical Sciences, Buffalo, NY

[4]Department of Medicine, F. Edward Hébert School of Medicine, Uniformed Services University of the Health Sciences, Bethesda, MD

[5]U.S. Army Futures Command, Combat Capabilities Development Command Soldier Center STTC, Orlando FL

[6]Centre for Systems Modelling and Quantitative Biomedicine, University of Birmingham, United Kingdom

[7]College of Engineering, Florida A&M University-Florida State University, Tallahassee, FL 32310, USA

**Abstract**

Objective motor skill assessment plays a critical role in fields such as surgery, where proficiency is vital for certification and patient safety. Existing assessment methods, however, rely heavily on subjective human judgment, which introduces bias and limits reproducibility. While recent efforts have leveraged kinematic data and neural imaging to provide more objective evaluations, these approaches often overlook the dynamic neural mechanisms that differentiate expert and novice performance. This study proposes a novel method for motor skill assessment based on dynamic directed functional connectivity (dFC) as a neural biomarker. By using electroencephalography (EEG) to capture brain dynamics and employing an attention-based Long Short-Term Memory (LSTM) model for non-linear Granger causality analysis, we compute dFC among key brain regions involved in psychomotor tasks. Coupled with hierarchical task analysis (HTA), our approach enables subtask-level evaluation of motor skills, offering detailed insights into neural coordination that underpins expert proficiency. A convolutional neural network (CNN) is then used to classify skill levels, achieving greater accuracy and specificity than established performance metrics in laparoscopic surgery. This methodology provides a reliable, objective framework for assessing motor skills, contributing to the development of tailored training protocols and enhancing the certification process.

# 1   Introduction

Motor skill assessment plays a vital role across diverse domains, including surgery[1], sports[2], and driving[3], serving as a benchmark for professional certification. However, prevailing methods for motor skill assessment predominantly rely on human administration[4], rendering them subjective and susceptible to biases, thereby undermining their reliability and reproducibility. Efforts have been made for objective[5] assessment by using kinematic measures, including tool motion[6], image and video of the performance[7][8][9], and brain activation[10][11]. Nevertheless, these methods overlook the intricacies of brain neural interaction mechanisms that differentiate experts from novices, limiting their interpretational scope.

Motor skill learning requires integrating information from disparate brain regions and their coordinated functioning[12]. The quantification of information transmission with directionality among the brain regions is known as directed functional connectivity (dFC), which profoundly influences task performance efficiency, rendering it a natural biomarker/metric for skill assessment. While previous research has explored cortical activation[13][14][10] as a biomarker of surgeons' skill levels, it often lacks insights into the underlying functional connectivity. Additionally, attempts to use connectivity for the objective assessment have been limited to univariate statistical methods[15][16] or exhibit limited interpretability[17][18] due to the use of undirected functional connectivity. Further, most studies historically considered the task as a whole[19][14], overlooking the dynamic and transient nature of connectivity during subtask executions, which hinders the identification of specific task elements crucial for skill assessment. While sliding window methods can capture dynamics connectivity[20][21][22], they cannot account for individual differences in task completion time, which may lead to different subtasks being compared within the same window[23].

To address these limitations, we propose the use of dFC combined with hierarchical task analysis (HTA) for detailed and objective skill assessment. HTA allows deconstructing complex tasks into more manageable subtasks[24], facilitating task standardization[25] thus enabling direct expert and novice comparisons and offering a detailed, equitable, and accurate evaluation. Analyzing dFC at the subtask level provides specific neurophysiological differences, aiding in the development of targeted training protocols for novices[26][27] to enhance their skill acquisition. We employ electroencephalography (EEG) to capture subtle, transient brain dynamics with millisecond-level precision[28] that provides direct insight into neural interaction mechanisms. We develop an attention-based LSTM model to compute non-linear Granger causality(nGC) as a measure of brain connectivity[29] between regions associated with psychomotor tasks. The brain regions analyzed include the prefrontal cortex (PFC), the primary motor cortex (M1), and the supplementary motor area (SMA). Given skill level-related differences in dFC, a 1D convolution neural network (1DCNN) is employed to evaluate skill level using the connectivity data. Our study marks a significant advancement in the field of motor skill assessment, with a particular focus on complex bimanual tasks such as those encountered in laparoscopic surgery. Figure 1 illustrates the overview of our study. Significantly, our approach, underpinned by deep learning, demonstrates that as a neural biomarker, dynamic dFC correlates with varying levels of surgical expertise. This innovative methodology not only enhances the specificity in distinguishing surgical motor skills but also proves to be more accurate than the current metrics used for certification in general surgery. By providing a practical, quantitative method to evaluate complex bimanual tasks, especially in high-stakes environments like surgery, this neural biomarker as an assessment metric contributes to the development of targeted training protocols and ensures both professional competence and patient safety.

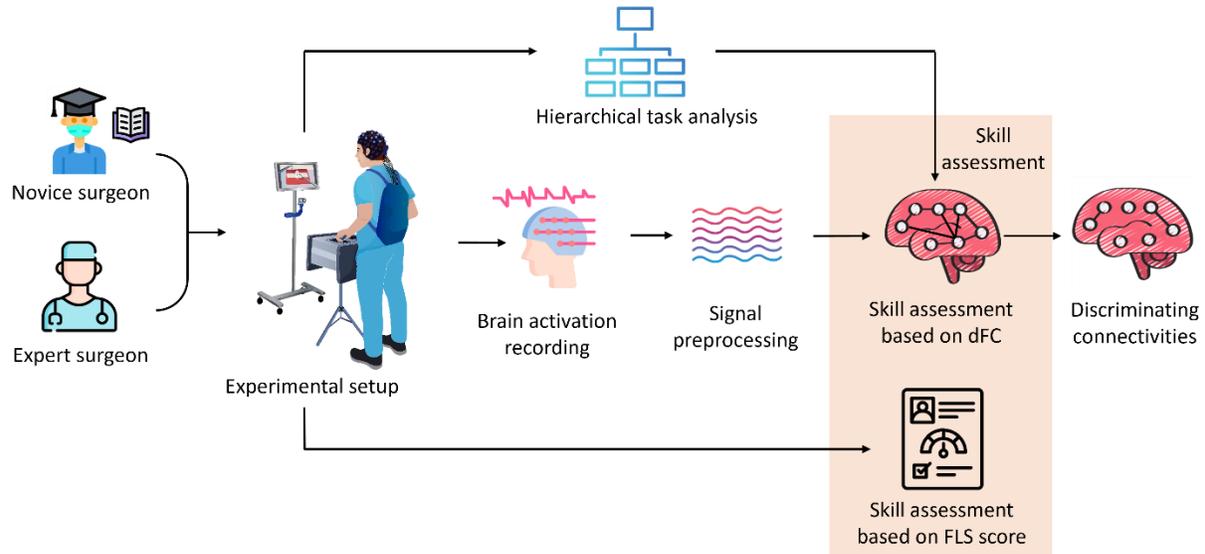

**Figure 1:** Overview of the study. Expert and novice surgeons perform the suturing with intracorporeal knot tying task of the Fundamentals of Laparoscopic Surgery (FLS) skills assessment system. Concurrently, their brain activations are recorded which are preprocessed and analyzed offline. Hierarchical Task Analysis (HTA)[24] is used to segment the task into several subtasks, and dynamic functional connectivity (dFC) is computed at each subtask using an LSTM model. Skill assessment is performed at each subtask using 1D CNN model. To validate the effectiveness of our approach, the dFC-based skill assessment is compared with the established FLS score-based assessment. This comparison provides a benchmark for accuracy in evaluating surgical proficiency. Recursive feature elimination (RFE) is applied to identify key discriminating connectivities between expert and novice surgeons for neurophysiological interpretation at the subtask level.

## 2 Results

### Task-specific skills assessment with dFC

To demonstrate the potential of dFC as a quantitative measure for evaluating bimanual skill proficiency, we selected a bi-manual suturing with intracorporeal knot-tying task, which is one of the five manual skills tasks included in the Fundamentals of Laparoscopic Surgery (FLS) program. Demonstrating proficiency on the FLS tasks is a pre-requisite for board certification in general surgery. Our study involved novice surgeons (1st- to 3rd-year surgical residents) and expert surgeons performing the suturing task in a standardized FLS box trainer.

Using hierarchical task analysis (HTA), the suturing task was divided into four coarse-level and thirteen fine-level subtasks by subject matter experts [30][31][32][33]. Figure 2 depicts these subtasks in procedural order (from top to bottom). dFC was computed using the non-linear GC method and skill assessment based on dFC was performed for each subtask. We explored multiple machine learning techniques and, after 5-fold stratified cross-validation, found that the 1D convolutional neural network (1DCNN) consistently outperformed other methods, leading us to select it for subsequent analyses, as detailed in the supplementary materials (S. Figure 1).

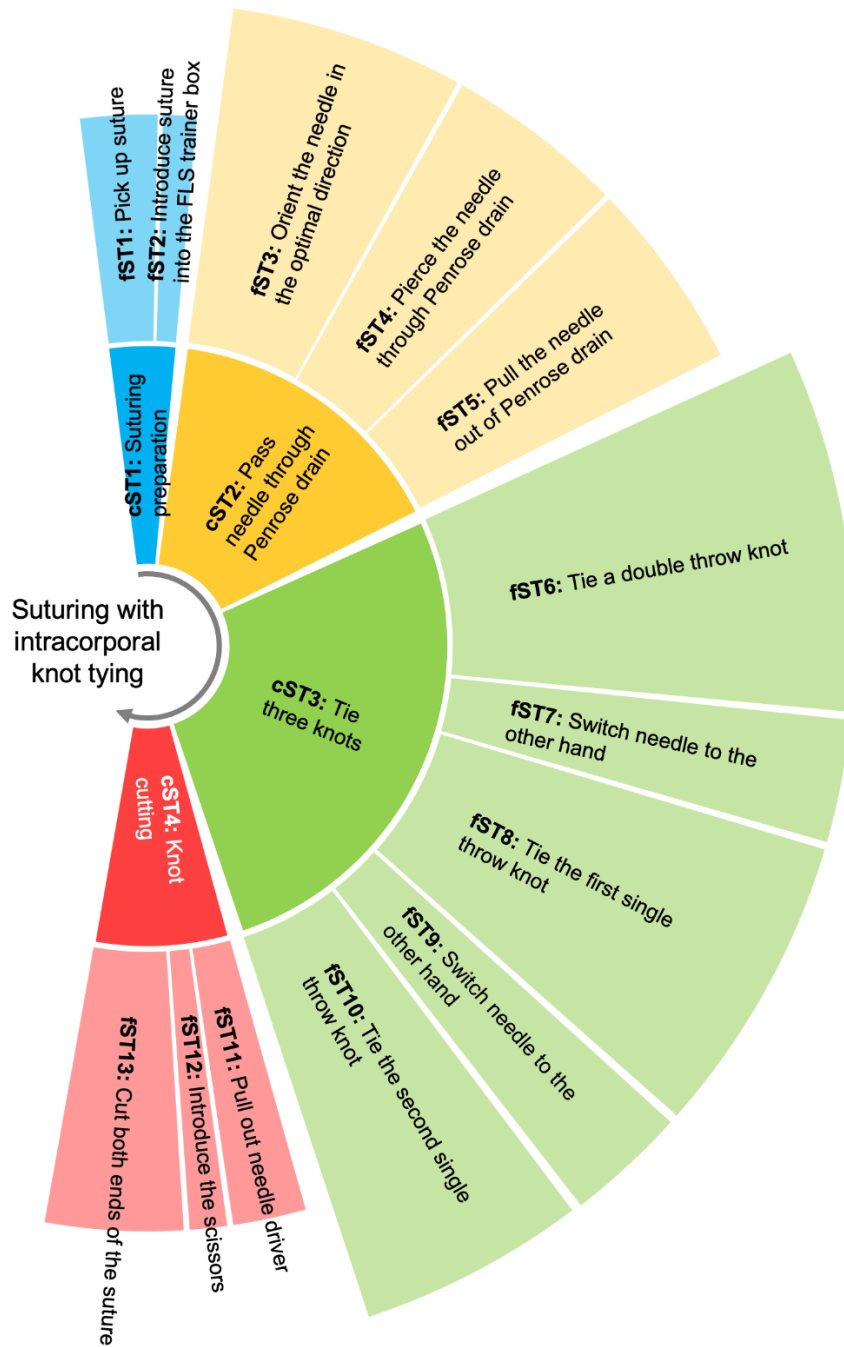

**Figure 2:** Hierarchical task analysis of the FLS suturing task with intracorporal knot tying. The task is divided into four coarse-level subtasks (cST), starting with picking up the suture and ending with knot cutting, that are further divided into thirteen fine-level subtasks (fST). The flow of action is indicated by an arrow.

In this study, rather than setting an arbitrary accuracy threshold to determine significantly differing skill levels [34][35], we consider a subtask to be significantly differentiating if its accuracy is greater than FLS score-based skill assessment. The FLS score based assessment achieved an accuracy of 82.8% (Table 3) which is further described in the "Skill assessment with established FLS performance score" section below. Skill difference was observed for all of the coarse level subtasks (cST, see Table 1). The observed differences at the coarse level served as preliminary indicators in recognizing skill differences, prompting a deeper examination at a finer subtask (fST) level. A finer-level analysis offers a deeper insight into skill disparities and is crucial for delivering targeted feedback to novices, helping them enhance their task performance. The details of the analysis at each fine level subtasks are provided in Table 2. We found that the skill difference observed while preparing to suture ( cST 1) was driven by the difference in inserting the suture into the FLS trainer box (fST 2). The skill difference while passing the needle through the Penrose drain (i.e. cST 2) was driven by the skill difference while piercing the needle through the Penrose drain (fST 4) and pulling the needle out of the Penrose drain (fST 5). The skill difference while tying the three knots (cST 3) was indicative of skill difference at all of its child subtasks (i.e., from tying a double throw knot (fST 6) to tying the second single throw knot (fST 10). Similarly, the skill difference at knot cutting (cST 4) was indicative of skill difference at all of its child subtasks( i.e., from pulling out one of the needle drive from the FLS trainer box (fST 11) to cutting both ends of the suture (fST 13)).

In our study, we have defined the novice class as the negative class, meaning that specificity refers to the model's ability to correctly classify novices (true negatives), while sensitivity refers to its ability to correctly classify experts (true positives).The high specificities of our models in skill assessment (Tables 1 and 2), indicates that the model makes fewer mistakes in recognizing novices. This is crucial to avoid the false certification of novices, which is essential for ensuring patient safety. Further, the high sensitivity and Matthew Correlation Coefficient (MCC) indicate that the model is equally effective in recognizing experts, however in this study, we focus more on accuracy and specificity which are important to identify unqualified surgeons and avoid their false certification. The confusion matrices and receiver operating curve (ROC) for the coarse and fine level analysis are provided in supplementary materials (S. Figure 3-19).

We used a recursive feature elimination (RFE) [36][37] method at the fine level to identify brain connectivities with discriminating information between expert and novice. RFE is a feature selection technique that iteratively removes the least discriminating features to identify the most discriminative ones. It works by training a machine learning model (1D CNN in our study), ranking the features based on their importance for classification, and eliminating the least important ones, then retraining the model. This process continues until all features that did not impact the accuracy of the model are eliminated. The remaining features are identified as the features with the most discriminating information. Details of this method are further provided in the "Methods" section. The most discriminating connectivities for each fine level subtasks are shown in Figure 3. An indepth discussion on the neurophysiology of the dominant discriminating connectivites are discussed in the "Discussion" section. The LM1→RM1, LM1→LPFC and RPFC→LPFC are found to be the dominant discriminating connectivities and are consistent with those reported in the literature for bimanual motor skills[16][38].

| cST | Coarse-level subtask | Accuracy | Specificity | Sensitivity | MCC |
|---|---|---|---|---|---|
| 1 | Suturing preparation | **0.903** | 0.974 | 0.783 | 0.793 |
| 2 | Pass the needle through Penrose drain | **0.905** | 0.914 | 0.893 | 0.807 |
| 3 | Tie three knots | **0.947** | 0.909 | 1.000 | 0.899 |
| 4 | Knot cutting | **0.907** | 0.962 | 0.824 | 0.805 |

**Table 1**: dFC-based skill assessment for the coarse-level subtasks. MCC stands for Matthews Correlation Coefficient (MCC).

| fST | Fine-level subtask | Accuracy | Specificity | Sensitivity | MCC |
|---|---|---|---|---|---|
| 1 | Pick up suture | 0.823 | 0.696 | 0.942 | 0.663 |
| 2 | Introduce suture into the FLS trainer box | **0.879** | 0.973 | 0.759 | 0.762 |
| 3 | Orient the needle in the optimal direction | 0.807 | 0.852 | 0.767 | 0.619 |
| 4 | Piercing the needle through two marks on the Penrose drain | **0.851** | 0.917 | 0.774 | 0.702 |
| 5 | Pull the needle out of Penrose drain | **0.853** | 0.875 | 0.821 | 0.696 |
| 6 | Tie a double throw knot | **0.887** | 0.871 | 0.903 | 0.775 |

| 7 | Switch the needle to the opposite hand | **0.932** | 0.938 | 0.917 | 0.834 |
| 8 | Tie the first single throw knot | **0.957** | 0.933 | 1.000 | 0.914 |
| 9 | Switch the needle to the opposite hand | **0.905** | 0.926 | 0.867 | 0.793 |
| 10 | Tie the second single throw knot | **0.957** | 0.967 | 0.938 | 0.904 |
| 11 | Pull out one of the needle drivers | **0.878** | 0.909 | 0.842 | 0.755 |
| 12 | Introduce the scissors into the box | **0.891** | 0.926 | 0.842 | 0.775 |
| 13 | Cut both ends of the suture | **0.933** | 0.929 | 0.941 | 0.861 |

**Table 2**: dFC-based skill assessment for the fine-level subtasks.

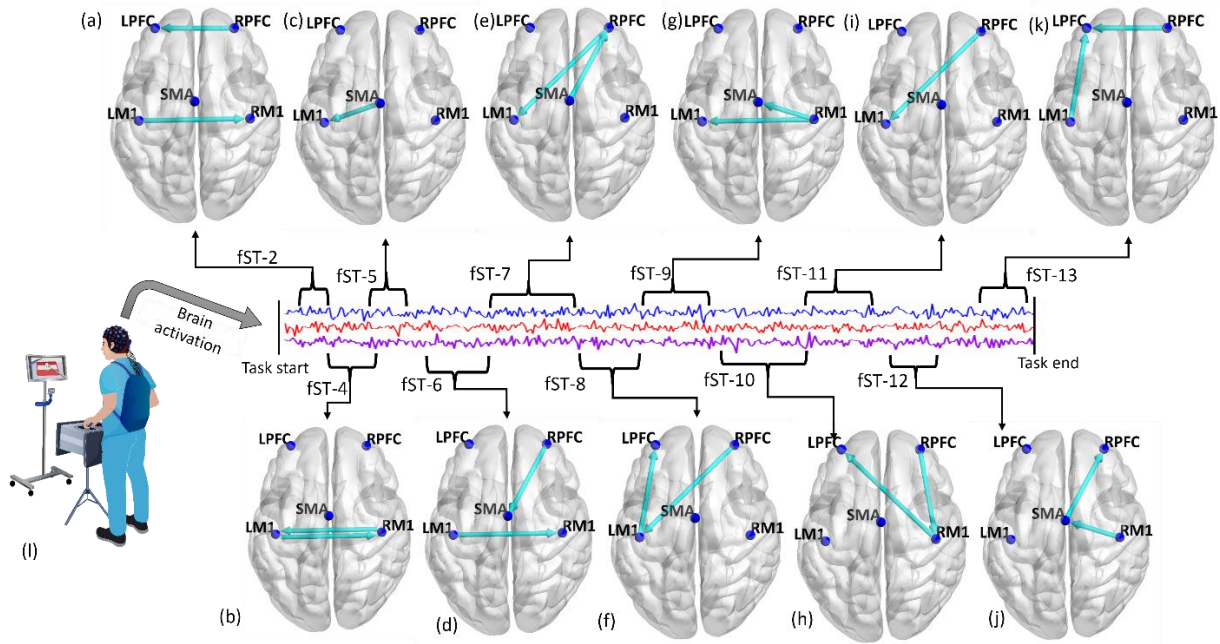

**Figure 3:** The upper and lower rows in the figure represent discriminating connectivities at various fine-level subtasks, while (a) introducing suture into the FLS trainer box (b) piercing the needle through two marks on the Penrose drain (c) pulling the needle out of Penrose drain (d) tying a double throw knot (e) switching the needle to opposite hand (f) tying the first single throw knot (g) switching the needle to the opposite hand (h) tying a the second single throw knot (i) pulling the needle-driver out of the trainer box (j) introducing scissor into the trainer box (k) cutting both ends of the suture (l) schematic depicting the FLS box simulator where surgeons perform the suturing and intracorporal task. An EEG system is used to measure functional brain activation in real time. Note: fST denotes fine subtask and arrow indicates the flow of information from source to target brain region.

## Skill assessment with established FLS performance score

Following task performance, we recorded FLS performance scores for each surgeon. The FLS scoring methodology was obtained from the FLS Committee through a nondisclosure agreement. Figure 4(a) reports the descriptive statistics of the FLS performance score for the novice and expert surgeons, where experts significantly outperformed novice surgeons ($p < 0.01$, $t= 12.620$, $df=44$, two-tailed t-Test).

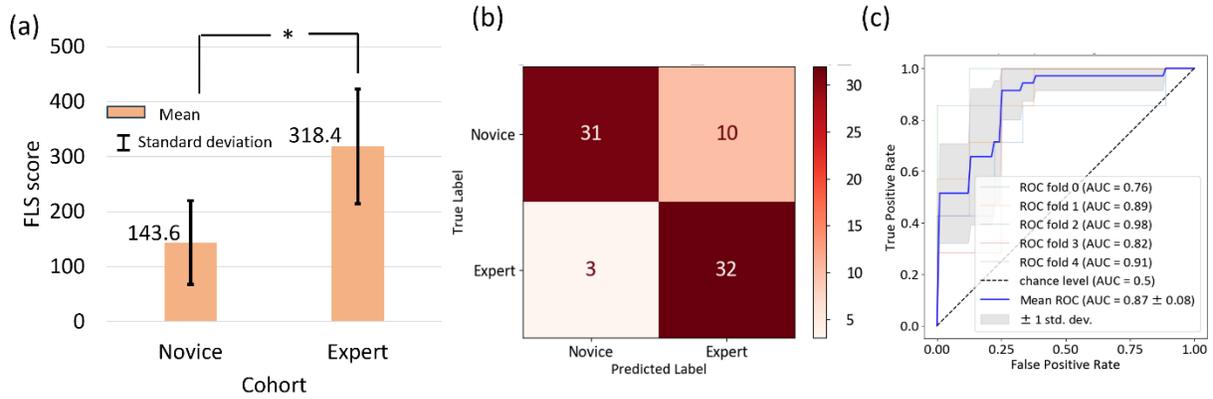

**Figure 4:** (a) FLS score of experts and novices and (b and c) Confusion matrix and ROC curve for skill assessment based on prevailing FLS score metric.

Figure 4(b-c) reports the confusion matrix and ROC curve of the kernel support vector machine (kSVM) classifier based on the FLS score. Note, that the highest classification accuracy with FLS score was obtained with kSVM compared to 1DCNN. The descriptive performance metrics of the classifier are provided in Table 3.

| Accuracy | Specificity | Sensitivity | MCC |
|---|---|---|---|
| 0.828 | 0.758 | 0.914 | 0.684 |

**Table 3:** FLS score-based skill assessment

The result indicates the poor performance of the FLS score-based assessment as it misclassified 10 novices as experts (see Figure 4(b)). Further, despite the 82.8% accuracy using the FLS score, relying solely on the FLS as a metric for skill assessment limits our understanding of expertise, as it lacks insights into underlying neurophysiology and does not provide information about the specific subtasks where skill difference exists. This underscores the importance of incorporating brain connectivity analysis with HTA for a comprehensive skill assessment.

## Brain-based versus FLS performance score

We compared the dFC-based classification with the standardized FLS score-based classification across all subtasks. Results indicate that classification is relatively poor when considering FLS performance scores only (specificity = 75.8%; see Figure 5(a) and 5(b) first element on the x-axis).

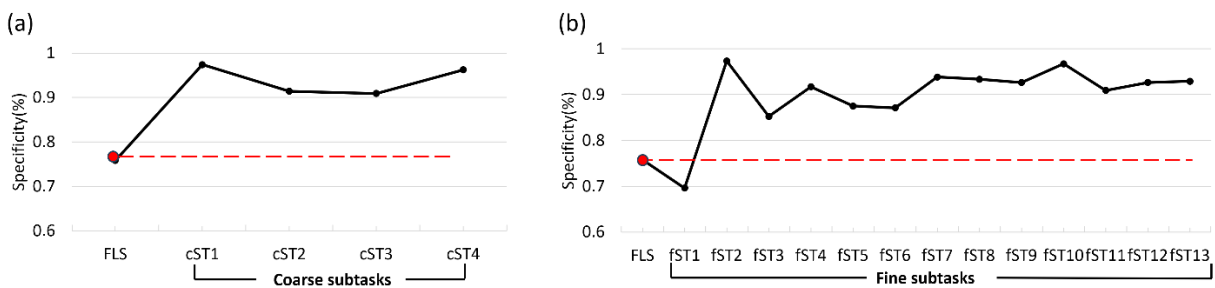

**Figure 5:** Comparison of specificity of skill assessment for coarse (a) and fine (b) level analysis with FLS score-based specificity. The FLS score-based results are highlighted in red dots.

On the other hand, the misclassification of novices is lower when using dFC, as demonstrated by higher specificity (see Figure 5(a) and (b)) in the coarse and fine analysis. The superiority of dFC is consistent across all coarse subtasks. A similar pattern is observed in the fine analysis (see Figure 5(b)), except for subtask 1, which involves picking up the suture from the table, likely due to the simplicity of this subtask. These results indicate that dFC is superior to FLS score-based assessment at recognizing the unqualified novice and avoiding false FLS certification.

# 3 Discussion

Our neural biomarker not only achieved higher classification accuracy than the traditional FLS score for skill assessment, but it also surpassed the other existing methods in the literature [39][40][41][42][43][44]. Achieving a maximum accuracy of 95.7% and specificity of 93.3% (Table 2, row 8), it surpassed the FLS score, which achieved an accuracy of 82.8% and a specificity of 75.8% (Table 3 and Figure 4). This advantage is observed across multiple coarse- and fine-level subtasks, as presented in Tables 1 and 2. The high specificity of our proposed deep learning-based model for most of the subtasks is critical in identifying unqualified surgeons and preventing their false certification. This ensures patient safety and upholds high healthcare standards, addressing human error—a leading cause of death in operating rooms[45]. The overall classification accuracy in these subtasks suggests that neurophysiological data can provide a highly accurate biomarker for distinguishing skill levels, most of the time more so than traditional FLS metrics. This could be due our proposed deep learning model's potential to capture subtle nuances in performance through brain connectivity that may not be as apparent through observational FLS scoring. The use of HTA provided us with the ability to pinpoint specific subtasks where novices require improvement. Granular neurofeedback can be useful in designing training programs tailored to improve novice skills[46][47]. The coarse level analyses were indicative of skill difference, and to delve deeper into a particular action/movement and its neurophysiology, we performed fine level analyses, which are discussed in the following paragraphs.

Motor action requires dynamic interaction of various brain regions involved in different aspects of movement preparation and execution. In fine subtask 2, which involves introducing the suture into the FLS trainer box, one of the discriminating connectivities is LM1→RM1 (Figure 3(a)). To understand this observation, it is essential to delve into the functions of these brain regions and how they relate to complex motor tasks. The primary motor cortices in both hemispheres, connected via the corpus callosum[48], enable interhemispheric communication crucial for interhemispheric motor coordination, inhibition[49], and hand-eye coordination[50]. Studies have demonstrated that interhemispheric connectivity in the motor cortex is associated with better motor coordination and performance[51]. During this subtask, the operator brings the suture close to the opening in the trainer box and guides it through the hole to the middle of the box. This task involves unilateral movement of the right hand while the left hand remains inactive. Executing strictly unilateral motor movements requires the suppression of the natural tendency toward bimanual synchronization[52][53] which requires interhemispheric interactions of homologous motor cortices[54][55][56]. Here, the inhibitory information is transferred from LM1 to RM1 to suppress left-hand movement. The process of unilateral movement suppression is known as transcallosal interhemispheric inhibition (IHI)[57][58] and the connectivity belongs to the "non-mirroring" transformation network[59]. This finding is in accordance with recent studies on interhemispheric interactions in which unilateral hand movements induced mechanisms that suppressed motor activation of the resting hand[58]. The other discriminating connectivity in this subtask is RPFC→LPFC (Figure 3(a)). The RPFC mediates spatial attention control and feedback monitoring[60] which are essential for the LPFC's task-set maintenance functions (planning and the cognitive control of motor actions[61][62]) to execute delicate movements in a 3D environment and guiding the suture through the hole. Experts likely have developed more efficient attentional and monitoring mechanisms, allowing them to maintain focus on the precise movements and trajectory required to bring the needle from the table and introduce it into the box through the small hole.

In fine subtask 4, while piercing the needle through the two dot marks on the Penrose drain, the discriminating connectivities are LM1→RM1 and RM1→LM1 (see Figure 3(b)). It is well known that the primary motor cortex (M1) is crucial for executing fine voluntary motor actions. The task of piercing the needle into the Penrose drain requires the operator to use both hands simultaneously—one to press and close the slit, and the other to orient and pierce the needle through the drain. The coordinated use of both hands and maintaining hand-eye coordination, requires efficient interhemispheric communication[63]. Further holding the needle in the correct orientation and applying piercing force are predominantly managed by the primary motor cortices[64]. Such bidirectional inter-hemispheric motor communication is facilitated by the corpus callosum, allowing for the integration and coordination of motor commands[48]. Studies have shown that expert performers in various motor domains often exhibit enhanced inter-hemispheric connectivity compared to novices[65].

In fine subtask 5, which involves pulling the needle out of the Penrose drain, SMA→ LM1 (Figure 3(c)) is the discriminating connectivity. The SMA is essential for preparing and initiating complex sequential motor actions, timing them accurately[66], and coordinating various motor movements[67][68][69]. This information is transferred to LM1 through the densely connected neurons[66] between SMA and LM1 regions. The LM1 sends the executive information (like motor angle and muscle forces) to the muscles via the corticospinal tract to execute the movement and control them. Thus, SMA→LM1 links cognition to movement[70][71]. Here, participants must carefully grasp the needle (protruding out of the Penrose drain) and pull it out ensuring enough suture is available on either side of the drain for the subsequent knot-tying steps, which involve a combination of motor preparation, timing, and controlled execution. Such sequential motor action is initiated in experts based on internal cues[72] (e.g., memory), however, the novice lacks such cues owing to inexperience. This finding is in line with a longitudinal sequential motor learning study that found attenuation of functional connectivity in premotor associative networks (SMA→LM1)with extended practices[73]. Further, this finding supports the facilitatory role of SMA to the hand area of M1[71] for organized sequential motor execution.

In fine subtask 6, which involves tying a double throw knot, one of the discriminating connectivities is RPFC→SMA (see Figure 3 (d)). RPFC is essential for cognitive control[62], strategic planning[61], and working memory[74][75], and plays a key role in conscious monitoring of actions[76]. In this subtask, participants must coordinate multiple elements: manage the needle and suture, plan the knotting steps, and monitor the execution. While performing these elements, the cues inside the trainer box continuously changes. In such a dynamic workspace/environment, the conscious monitoring of action is needed to adjust the action plan and update the working memories. Based on these action plans, the SMA coordinates the execution of muscles in the motor systems to perform movement. The information related to cognitive control, action monitoring, and working memory is transferred to the SMA to select the intended action[77] and coordinate the motor execution. Furthermore, wrapping the suture around the needle-driver to form the loop requires precise spatial planning and motor execution, and taking out the needle-driver from the loop involves careful timing and action monitoring. Both of these processes are mediated by the RPFC→SMA connectivity[78] which is in accord with the cognitive-motor integration model[79]. Another discriminating connectivity is LM1→RM1. The LM1 and RM1 execute the voluntary motor actions and govern coordinated in-phase or/and anti-phase bilateral movements of hands which are required in tying a double throw knot. This finding is substantiated by our previous study on skill level differentiation for the FLS pattern-cutting tasks[16].

While transferring the needle from one hand to another in fine subtask 7, the discriminating connectivities are SMA→RPFC, and RPFC→LM1 (see Figure 3(e)) between expert and novice surgeons. The SMA is essential for sequential motor tasks[80], coordinating bilateral movements[81], and is involved in visually guided movements/feedback[82] in the 3D environment[83]. In this subtask, the SMA likely plays a critical role in sequencing[84] the hand movements(like bringing both hands close, releasing the needle, and securely grasping with another hand) required to transfer the needle and send the visually guided feedback to the RPFC. The RPFC uses this feedback and information to formulate cognitive plans for motor execution (adjustment[85][86] and/or inhibition of unwanted hand movements[87]). To do this, the RPFC needs to constantly maintain spatial focus and execute the cognitive plan or make the adjustment, requiring it to recruit the attention network[88] i.e. RPFC→LM1. The RPFC→LM1 is a component of a frontoparietal network that not only maintains focus on the task being executed but also mediates a top-down control and execution of the cognitive plans. The discriminating top-down connectivity (i.e. RPFC→LM1) is elicited by visual cues[89] here likely by the position of two needle drivers. Overall, the stimulus-driven control of special attention and feedback-driven control of execution is not developed in novices resulting in the differences.

For fine subtask 8, which involves tying a single throw knot, one of the discriminating connectivities is RPFC→LM1 (see Figure 3(f)). A notable strategy employed by all operators was wrapping the thread around the needle driver, which demands precise spatial awareness and motor control. This action is likely facilitated by RPFC to LM1 connectivity. The connectivity from the RPFC (responsible for strategic motor planning, motor feedback monitoring, and adjustment[86]) to LM1 (responsible for fine motor control and execution[64]) reflects the translation of complex motor plans into execution, which is needed to create the loop and tie the knot. The other discriminating connectivity is the LM1→LPFC. The LM1 is crucial for encoding, consolidating, and retrieving motor memories essential for skilled movements[90][91]. LM1 also processes somatosensation relevant to the motor system[92][93]. Connectivity from LM1 to LPFC reflects the transmission of somesthesis and retrieval of learned skills to the prefrontal cortex for making decisions and plans for suture cutting. The complex sequence of opening the second needle-driver, holding the thread, and removing the needle-driver from the loop while keeping the thread in place involves efficient somesthesis and motor memory retrieval (in experts), potentially facilitated by LM1→LPFC connectivity. According to the dual-process theory, novices, lacking procedural knowledge and expertise, tend to engage in more deliberate, conscious processing[94][95] leading to the difference in connectivity against experts.

In fine subtask 9, while switching the needle to the opposite hand, RM1→LM1 (Figure 3(g)) is one of the discriminating connectivities. Studies indicate that efficient motor coordination relies on interactions between homologous motor areas across hemispheres[96] and is crucial for bimanual coordination. This subtask involves transferring the needle from one hand to the other, requiring effective interhemispheric communication for fine motor coordination between the hands likely facilitated by RM1→LM1 connectivity. The second discriminating connectivity is RM1→SMA. RM1 receives sensory inputs from the somatosensory region[97] and plays a role in motor memory retrieval. The sensory and motor memory information needs to be transferred to the SMA for coordinating, timing, and sequencing the motor actions to transfer the needle from one hand to the other[98][84]. Differences in connectivities reflect the superior timing and coordination abilities of experts compared to novices.

In fine subtask 10, while tying the second single throw knot, one of the discriminating connectivities is the RPFC→RM1(see Figure 3(h)). The RPFC is involved in spatial attention, motor monitoring, and strategic motor planning, and supports efficient motor performance by enabling real-time adjustments based on feedback [99]. Pulling both ends of the thread to tighten the knot involves the coordination of spatial attention, motor monitoring, and execution, likely supported by RPFC→RM1 connectivity. The feedback and adjustments are essential for achieving goal-directed motor behavior flexibly and adaptively to maintain the loop on a low-friction needle-driver stem and tie the knot securely. The observed difference could be due to novices showing less efficient connectivity, indicating a more effortful and less coordinated approach to movement monitoring and adjustment[100]. The other differentiating connectivity is the RM1→LPFC, which is a component of the frontoparietal network[101][102]. The frontoparietal network is activated in learning long motor sequence tasks [103]as it requires effective coordination between motor and cognitive regions[104][105]. This finding aligns with the notion that skill acquisition improves the integration and communication in the frontoparietal network[104].

In fine subtask 11, which involves pulling the needle-driver out of the FLS trainer box, the RPFC→LM1 (see Figure 3(i)) connectivity was notably discriminating. To effectively pull the needle-driver through the small opening, precise spatial attention, plan, and motor control are essential to prevent obstruction and ensure perpendicular alignment to the hole in the trainer box. This task relies on the RPFC's ability to oversee spatial aspects and monitor motor actions[99] as discussed above, with these inputs translated into precise movements by the LM1. Experts likely demonstrate more efficient RPFC→LM1 connectivity, reflecting superior spatial awareness and adjustment capabilities. Experts often use advanced techniques, such as grabbing both ends of the suture with the non-dominant hand to facilitate subsequent tasks like suture cutting. This skill showcases enhanced motor planning and execution, necessitating the recruitment of RPFC→LM1 connectivity.

In fine subtask 12, which involves introducing the scissors into the FLS trainer box, one of the discriminating connectivity is RM1→SMA (see Figure 3(j)). The RM1 is critical for executing skilled movements and integrating sensory feedback into the SMA[106]. Guiding the scissors through the small opening and adjusting its orientation necessitates precise motor execution and continuous motor sensory feedback to SMA for coordination, supported by RM1→SMA connectivity. The other discriminating connectivity is SMA→RPFC. The task of inserting the scissors requires a series of sequential actions: orienting the scissors correctly, pulling them toward the operator, adjusting their orientation, and guiding them through a small opening. The SMA not only coordinates[107] these tasks but sends information to the RPFC for proper spatial attention[89], adjusting orientation[108] or inhibition of inappropriate actions[109]. The experts tend to have better integration of feedback with sequential movement generation compared to noivices, enabling more precise adjustments[110] [111].

In fine subtask 13, which involves cutting both ends of the suture, RPFC→LPFC is one of the discriminating connectivities (Figure 3(k)). Cutting the suture accurately involves strategic planning (e.g., cutting both ends of the suture at once or separately), with connectivity from RPFC→LPFC indicating integration of spatial attention and motor monitoring (functions of the RPFC) with motor planning and organization (functions of the LPFC). Locating the suture and orienting the scissors demands precise spatial attention and motor planning, supported by the RPFC→LPFC connectivity. Further, the action selection when presented with multiple options is controlled by interhemispheric prefrontal connectivity[112] could be mediating the difference between experts and novices. Another differentiating connectivity is the LM1→ LPFC. The LM1 is crucial for execution and motor memory retrieval essential for skilled movements[90][91]. LM1 processes somatosensation relevant to the motor system[92][93].

Connectivity from LM1 to LPFC reflects the transmission of somesthesis and retrieval of learned skills to the prefrontal cortex for making decisions and plans for suture cutting. Both of these connectivities are in accordance with our previous study on expert versus novice for the FLS pattern cutting task which uses scissors to cut a circular mark on a gauge[38]. Experts likely demonstrated advanced motor planning, sensory information integration, and execution capabilities allowing for more efficient and accurate cutting of both ends of the suture compared to novice.

While this study provides valuable neural biomarker for skill assessment with insights into the neurophysiological differences between experts and novices at the subtask level, several limitations should be acknowledged. The sample size of both groups is relatively small, and future studies with larger cohorts are needed to validate the findings. The study focused specifically on a basic surgical skill from a manual skills exam. Examining a broader range of fields could provide efficacy of the proposed biomarker. Investigations into the extended longitudinal development of these neurophysiological and behavioral differences could provide additional insights into the progression of surgical skill acquisition along with the skill assessment.

# 4 Material and methods

## Hardware and equipment

We used a 32-channel wireless LiveAmp system (Brain Vision, USA) EEG montage to record the brain activity of the surgeons. EEG is a neuroimaging tool with high temporal resolution, non-invasiveness, and portability. The brain signals were obtained at a sampling frequency of 500 Hz using active gel electrodes. The electrode placement can be seen in the montage in Figure 5 and S. Figure 2. The lead field of the EEG montage is shown in supplementary materials (S. Figure 20). The electrode nomenclature follows the standard 10-5 system[113]. Odd numbers in Figure 6 represent electrodes on the left hemisphere, and even numbers represent electrodes on the right hemisphere. EEG operates by continuously monitoring the postsynaptic potentials generated by millions of neurons, offering millisecond-level resolution and direct insight into neural circuit operations. Synchronous neuronal activity generates electrical waves detectable by scalp electrodes, forming rhythmic EEG patterns or oscillations. The oscillations can be categorized into delta (δ; 0.5–3 Hz), theta (θ; 4–7 Hz), alpha (α; 8–13 Hz), beta (β; 14–30 Hz), and gamma (γ; 30–50 Hz) frequencies. Notably, alpha oscillations, linked to both cognitive[114] and motor processes[115][116] [117][118], were utilized in this study. They play a key role in procedural motor learning and coordination.

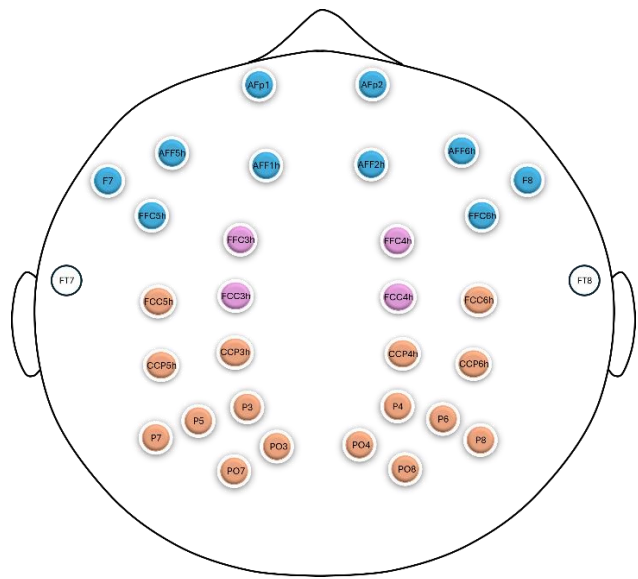

**Figure 6:** EEG Montage used to measure brain activation. Thirty-two electrodes were used at the predetermined location on the brain to cover the prefrontal cortex, supplementary motor area, and primary motor cortex.

## Participants and experimental procedure

Fifteen healthy medical residents (novices) and fifteen attending surgeons (experts) were recruited to participate in this study, which was approved by the Institutional Review Board at the University at Buffalo, NY. The participants performed the suturing with intracorporeal knot tying task, which is one of the five manual skills tasks of the FLS educational program, proficiency in which is a prerequisite for board certification in general surgery. The participants' demographics are provided in supplementary materials (S. Table 1). We defined surgical novices and experts based on their years post graduation and the number of laparoscopic procedures completed according to the literature[119]. All groups were independent; i.e., each subject belonged to only one group. Verbal instructions were provided to ensure that the participants were aware of the task steps[120]. The task began with a 120-second resting period, followed by a maximum of 600 seconds allocated for task performance. All subjects completed at least three trials. Note that a few subjects skipped some of the subtasks despite being instructed, and three experts had recording issues for EEG and were excluded from the analysis. Such subjects were removed from the pertaining subtask analysis. The EEG montage was carefully placed on the scalp by aligning the CZ, FP1, and FP2 landmarks on the head and the marked landmarks on the cap. While the subject performed the FLS suturing task, continuous EEG recording captured brain activation. All environmental distractions were removed from the surroundings during the resting and task periods; see Figure 3(I) for environmental setup. The FLS score was computed for each trial by a proctor based on completion time and errors [121], with a higher score indicating better performance.

## Preprocessing

The recorded EEG signals were preprocessed and analyzed offline. The open-source EEGLab toolbox [51] (https://sccn.ucsd.edu/eeglab/index.php), which is implemented in MATLAB, was used to preprocess the raw brain activation signals to remove the artifacts. The signals were first down-sampled to 250Hz, and a high pass filter at 1Hz was applied to remove linear trends or signal drift, and electric interference was removed at 60Hz. Three approaches were used to eliminate bad channels from the data. Firstly, flat channels were removed. Secondly, channels with significant noise were identified and removed based on their standard deviation. Lastly, channels that exhibit poor correlation with other channels were removed using a rejection threshold set at 0.8 for channel correlation. The channels, if removed, were interpolated using neighboring channels' information and spherical spline interpolation. The average reference was computed by subtracting the average of all electrodes from each channel. The current source density (CSD) of the cortical oscillators was computed using eLORETA software.

The 30 channels were grouped into 5 distinct brain regions of interest according to the anatomical structures as follows: the left prefrontal cortex (LPFC channels: AFp1, AFF5h, F7, AFF1h, and FFC5h), right prefrontal cortex (RPFC channels: AFp2, AFF6H, F8, AFF2h, and FFC6h), supplementary motor area (SMA channels: FFC3h, FCC3h, FFC4h, and FCC4h), left primary motor cortex (LM1 channels: FCC5h, CCP5h, CCP3h, P3, P5, PO7, P7, and PO3), and right primary motor cortex (RM1 channels: FCC6h, CCP6h, CCP4h, P4, P6, PO8, P8, and PO4) as they are associated with psychomotor skill learning[122][15]. All the possible pairs of connectivity were selected from LPFC, RPFC, SMA, LM1, and RM1 regions, resulting in twenty connectivity pairs (see S. Table. 2) which are used as features for the deep learning model.

## Directed functional connectivity via non-linear Granger causality:

In the linear Granger causality framework, a signal $X_i$ is considered "Granger causal" to another signal $X_j$, i.e., information flow directed from $X_i$ to $X_j$, if the inclusion of past values of $X_i$ improves the prediction of future values of $X_j$ beyond what is possible using only the past values of $X_j$. Given two stochastic signals $X_i(t)$ and $X_j(t)$ which are assumed to be jointly stationary, their autoregressive representations are described as

$$X_i(t) = \sum_{n=1}^{N} a_n X_i(t-n) + \epsilon_i(t) \qquad (1)$$

$$X_j(t) = \sum_{n=1}^{N} d_n X_j(t-n) + \epsilon_j(t) \qquad (2)$$

their joint representations are described as:

$$X_i(t) = \sum_{n=1}^{N} a_n X_i(t-n) + \sum_{n=1}^{N} b_n X_j(t-n) + \eta_i(t) \quad (3)$$

$$X_j(t) = \sum_{n=1}^{N} c_n X_i(t-n) + \sum_{n=1}^{N} d_n X_j(t-n) + \eta_j(t) \quad (4)$$

where $t = 0, 1, ..., M$ are the timestamps, N is the total lag, the noise terms $\epsilon_i$ and $\eta_i$ are uncorrelated over time with a mean of zero and a variance of $\sigma_{\epsilon i}^2$ and $\sigma_{\eta i}^2$, respectively. From equations 1 and 3, if $\sigma_{\eta i}^2 < \sigma_{\epsilon i}^2$, then $X_j$ is claimed to have a causal influence on $X_i$ [123][124]. In this case, equation 3 is better than equation 1 for predicting $X_i$(t). Otherwise, $X_j$ does not have a causal influence on $X_i$. Such causal influence, known as Granger causality (GC)[29][125][126], can be defined by

$$F_{X_j \to X_i} = \ln\left(\frac{\sigma_{\epsilon i}^2}{\sigma_{\eta i}^2}\right)$$

Thus $F_{X_j \to X_i} > 0$ indicates the causal influence from $X_j \to X_i$, and $F_{X_j \to X_i} = 0$, indicates no causal influence. Similarly, the causal influence from $X_i \to X_j$ is defined by

$$F_{X_i \to X_j} = \ln\left(\frac{\sigma_{\epsilon j}^2}{\sigma_{\eta j}^2}\right)$$

A conditional Granger causality[29][125] is defined for evaluating whether the interaction between two signals is direct or mediated by another signal as

$$F_{X_k \to X_i | X_j} = \ln\left(\frac{\sigma_{\epsilon k}^2}{\sigma_{\eta k}^2}\right)$$

where $\sigma_{\epsilon k}^2$ and $\sigma_{\eta k}^2$ are variances of noise terms, $\epsilon k$, and $\eta k$, respectively, of the following joint autoregressive representations:

$$X_i(t) = \sum_{n=1}^{N} a_n X_i(t-n) + \sum_{n=1}^{N} b_n X_j(t-n) + \epsilon_k(t) \quad (5)$$

$$X_i(t) = \sum_{n=1}^{N} c_n X_i(t-n) + \sum_{n=1}^{N} d_n X_j(t-n) + \sum_{n=1}^{N} e_n X_k(t-n) + \eta_k(t) \quad (6)$$

When there is a direct interaction from $X_k$ to $X_i$, the combined past measurements of $X_i, X_j, X_k$ result in better prediction of $X_i$.

The main limitation of this approach is the use of the AR model[127] used to describe interactions between signals which cannot capture non-linear dependencies [128][129]. Tank et al. (2021)[127][130] generalized the concept of Granger causality for nonlinear autoregressive models as follows:

$$X_i(t) = g_i(X_{<ti}, ..., X_{<tp}) + \eta_p(t)$$

where, $X_{<ti} = (X_i(t-N), ..., X_i(t-1))$ represents the history of signal $i$ and $g_i$ is a non-linear function mapping the lagged value of other $p$ signals to signal $i$. Following the Granger causality framework from equations 5 and 6, the non-linear autoregressive representations becomes

$$X_i(t) = g_i(X_{<ti}, ..., X_{<tp}) + \epsilon_k(t) \quad (7)$$

$$X_i(t) = g_i(X_{<ti}, ..., X_{<tk} ..., X_{<tp}) + \eta_k(t) \quad (8)$$

The causal influence is then defined by:

$$F_{X_k \to X_i | X_j} = \ln\left(\frac{\sigma_{\epsilon k}^2}{\sigma_{\eta k}^2}\right) \tag{9}$$

In this context, $F_{X_k \to X_i | X_j} = 0$ for $\sigma_{\epsilon k}^2 = \sigma_{\eta k}^2$ and $F_{X_k \to X_i | X_j} > 0$ for $\sigma_{\epsilon k}^2 > \sigma_{\eta k}^2$. Note that $\sigma_{\epsilon k}^2 \geq \sigma_{\eta k}^2$ always holds because the model generally improves with the inclusion of additional signals. Additionally, there is no upper bound for Granger causality, and it depends on the information added by other signals.

Recently, deep learning has been used to model the non-linear dependencies[131][132] within the Granger causality framework. In this study, we developed and implemented a novel approach to non-linear Granger causality analysis using an attention-based Long Short Term Memory (LSTM) model[133][134]. Unlike the traditional VAR model, the LSTM model can capture non-linear dependencies. Our model architecture features an encoder-decoder structure, harnessing the capabilities of the LSTM network to capture long-term dependencies inherent in signals. The encoder processes the input sequence to produce an encoded representation, while an attention mechanism calculates the importance of each time step within this sequence. This attention mechanism is implemented through a dense layer followed by a softmax activation which assigns dynamic weights to the encoded sequence. This enhances the model's ability to focus on significant time steps. The resulting weighted sequence is aggregated into a context vector, which encapsulates the most relevant temporal information. The decoder leverages this context vector, repeating it across each time step of the input sequence and initializing the LSTM layer with the encoder's final states. This approach allows the decoder to generate predictions that reflect the learned temporal patterns. Additive attention[135] was used in the encoder layer and the dot-product attention[136] layer is used in the decoder layer.

The model is trained using the mean squared error (MSE) loss function and Adam optimizer along with the early stopping option to prevent overfitting and ensure robust model performance. To rigorously evaluate the model's predictive accuracy, we utilized a suite of performance metrics, including mean absolute error (MAE), mean squared error (MSE), root mean squared error (RMSE), and R-squared ($R^2$) value. These metrics provide comprehensive insights into the model's ability to capture and predict complex temporal dynamics. Additionally, we analyzed the distribution and autocorrelation of residuals to assess the model's predictive reliability and to uncover potential temporal dependencies within the residuals. This evaluation framework ensures a robust model performance.

In our model, we train a LSTM encoder-decoder model on the past value of all 30 signals and predict the future of all the signals. The number of past timesteps of the signals is determined through exhaustive search as described below. This step uses the training and validation dataset. Once the model is trained, we construct full- and reduced-model datasets out of the test dataset. The full-model dataset includes all the signals. The reduced-model dataset is constructed by replacing the corresponding value of the signals with zeros. The future values of all the full- and reduced-model datasets are predicted by the trained LSTM model. The variance of the residual of the full- and reduced-model datasets is used to determine non-linear causality as defined in equation 9. Note that reduced-model predictions are associated with the corresponding removed signal. We use 80% of the data to train 10% to validate the full-model and the last 10% to test the reduced model. Optimal window size (corresponds to the number of lag terms in the AR process) is exhaustively searched from the following set [3,5,7,9] and kept the same for input and output windows. Sliding window of step size 1 and size of 5 was found to be the optimal for most of the subjects. Note that, in our method, a single LSTM model is trained, unlike others[138][139][140] where several models corresponding to reduced-models are trained separately. This increases the computational cost making them infeasible, especially for high temporal resolution signals like EEG. Some deep learning frameworks[127][132][140] rely on the interpretation of trained model weights for causal inference which is not reliable due to the random initialization of model weights and non-linearity of the models[141][142].

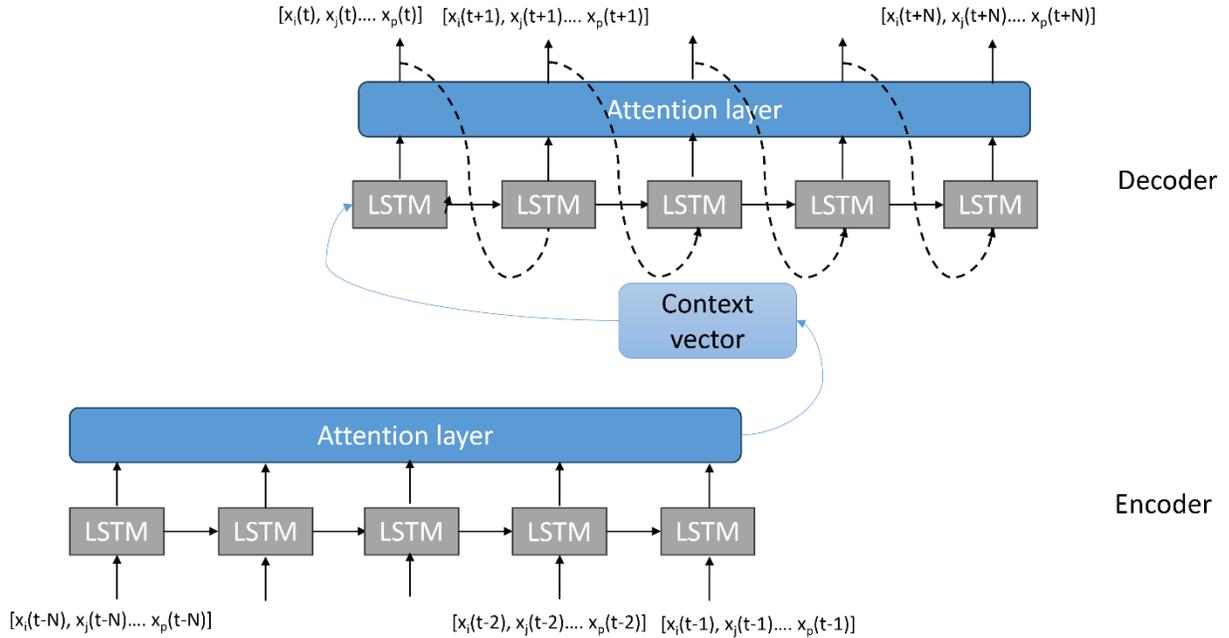

**Figure 7**: LSTM model architecture

To evaluate the efficacy of our model, we conducted experiments using two synthetic datasets for linear and non-linear systems. The results and figures of full and reduced model performance, along with the residual distribution and autocorrelation, are presented in the supplementary materials ("non-linear GC" section). Of note, the nGC was able to capture the causality for linear and non-linear systems, whereas the linear GC failed at non-linear systems.

In our study, each brain region contains several channels that measure brain activations. A common approach is to average the activation of these channels to get a representative activation of each brain region and subsequently compute regional brain connectivity[143][144][145]. However, averaging tends to smooth out the differences between channels, potentially losing unique information in each channel. To eliminate this limitation, we train the full model with all the channels. For reduced-model, we remove all the channels belonging to a particular region for which nGC needs to be computed. The noise variance of these channels is summed to get the total variance for this region and subsequently compute the nGC as described above. This is repeated for each brain region. By doing so, we ensure that the model captures the information of each channel.

## Task performance metrics

FLS scores for the suturing task were computed using a proprietary metric based on both time and error. Descriptive statistical analyses were conducted using IBM SPSS v26 software. Two-sample t-tests were utilized to identify statistically significant differences in FLS scores between the two groups. A confidence level of 99% was chosen as the minimum threshold for rejecting the null hypothesis.

## 1D CNN with recursive feature elimination (RFE)

For nGC-based skill assessment, we applied a 1D CNN (S. Figure 25) for the binary classification of experts and novices. The input to the model is a set of twenty connectivities obtained from the LSTM-based nGC model. The 1DCNN architecture has a Conv1D layer with 64 filters and a kernel size of 14, followed by a second Conv1D layer with 64 filters and a smaller kernel size of 4. A third Conv1D layer with 32 filters and a kernel size of 3 is then employed. To prevent overfitting, a dropout layer with a rate of 0.4 is applied after the first and second Conv1D layers. Next, a global average pooling layer is applied, followed by two fully connected dense layers with 64 and 32 units, respectively. All layers used the ReLU activation function to introduce non-linearity, except for the final output layer, which is a dense layer with a single unit and a sigmoid activation function that produces a probability value for binary classification. The model is trained with the Adam optimizer with a learning rate of 0.0001 and a binary cross-entropy loss function, with accuracy as the primary evaluation metric. Fivefold stratified cross-validation was applied to the training data to measure the model's performance, and here, we report the performance on the test sets.

To identify discriminating connectivities between expert and novice, we used the recursive feature elimination (RFE) approach[36][37] for each fine subtask separately. Starting with all features (i.e., 20 connectivities), it iteratively trains the 1D CNN model and evaluates its performance. At each iteration, the least important feature, as determined by the impact of its removal on the model's accuracy, is removed, and a new feature set is constructed for subsequent iteration. This process continues until all the features that didn't impact the accuracy of the model are eliminated. The remaining features are identified as the features with the most discriminating information. A list of dFC used as features for these models is shown in S. Table 2.

## 5 Acknowledgment

The authors gratefully acknowledge the support of this work through the Medical Technology Enterprise Consortium (MTEC) award #W81XWH2090019, and the U.S. Army Futures Command, Combat Capabilities Development Command Soldier Center STTC cooperative research agreement # W912CG2120001.

## 6 Data availability

Data used in this study may be available by contacting the corresponding author and codes can be found on GitHub (https://github.com/anilkamat/dFC_suturing.git.)

## 7 Conflict of interest statement

The authors affirm that there are no competing conflicts of interest present in their involvement with the subject matter or the publication of this work.

# 9 Supplementary materials

(a)

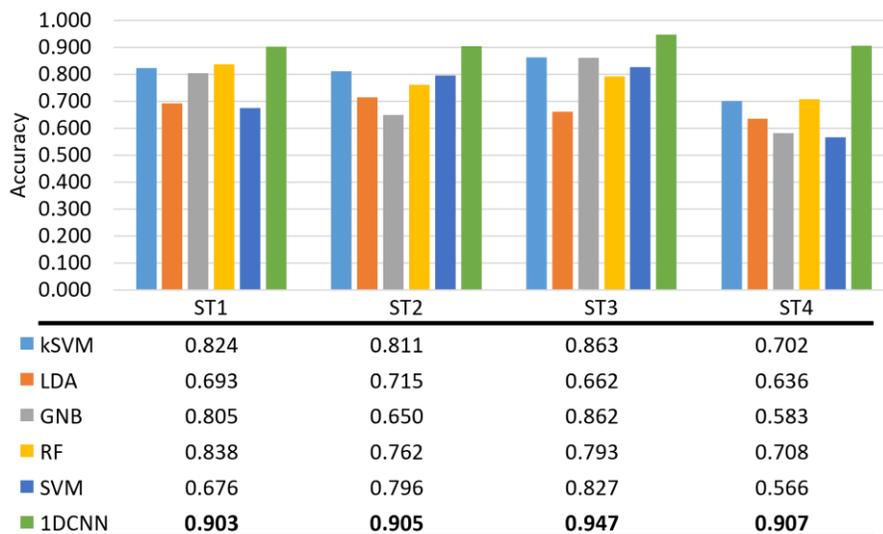

| | ST1 | ST2 | ST3 | ST4 |
|---|---|---|---|---|
| kSVM | 0.824 | 0.811 | 0.863 | 0.702 |
| LDA | 0.693 | 0.715 | 0.662 | 0.636 |
| GNB | 0.805 | 0.650 | 0.862 | 0.583 |
| RF | 0.838 | 0.762 | 0.793 | 0.708 |
| SVM | 0.676 | 0.796 | 0.827 | 0.566 |
| 1DCNN | **0.903** | **0.905** | **0.947** | **0.907** |

(b)

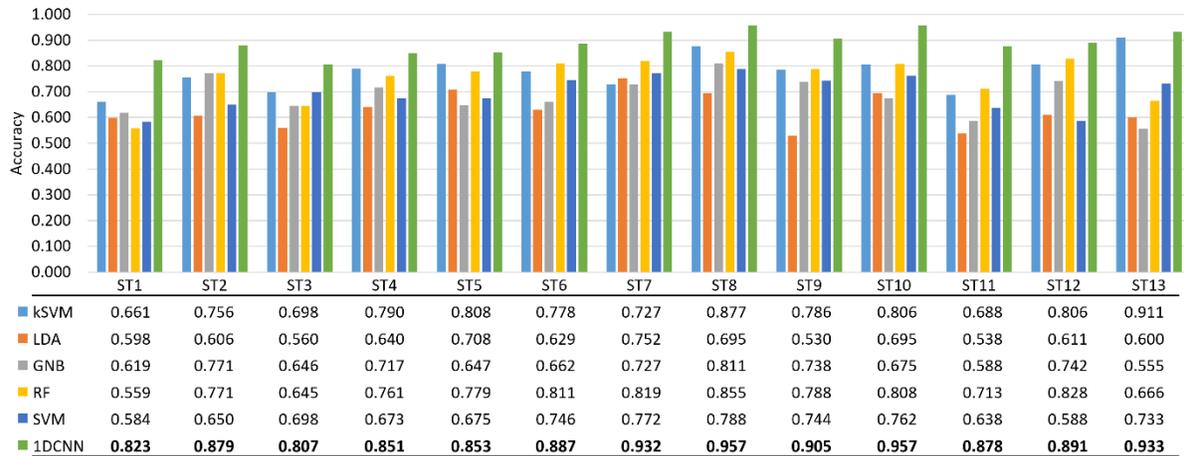

**S. Figure 1:** Performance of several machine learning models for surgical skill assessment based on dFC (a) coarse level (b) fine level. Notably, 1D CNN consistently outperforms other machine learning models.

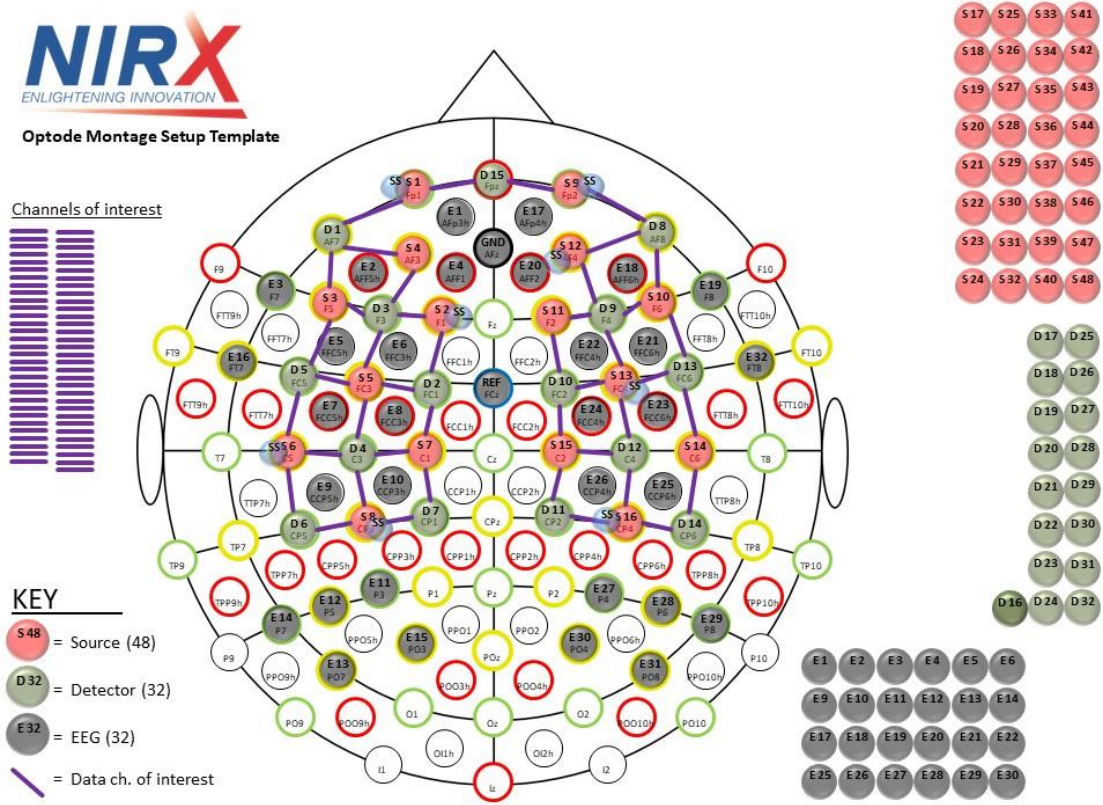

**S. Figure 2:** Montage used in measuring brain activation

**S. Table 1:** Demography of subjects

| Participant | Age | Gender | Dominant Hand | Med Student? | Year in Med School/Residency | Experience with laparoscopic tools? | Year of experience lap | FLS/FRS experience | Years of FLS/FRS |
|---|---|---|---|---|---|---|---|---|---|
| E01 | 56 | M | R | N | NA | Y | 25+ | Y | ~10 |
| E02 | 37 | M | R | N | NA | Y | 5 | Y | 3 |
| N01 | 26 | M | R | Y | M3 | N | NA | N | NA |
| N02 | 24 | F | R | Y | M2 | N | NA | N | NA |
| N03 | 22 | F | R | Y | M1 | N | NA | N | NA |
| N04 | 28 | M | R | Y | M1 | N | NA | N | NA |
| E03 | 22 | F | R | Y | M1 | Y | 1 | Y | 1 |
| N05 | 27 | M | R | Y | M4 | N | NA | N | NA |
| N06 | 26 | M | R | Y | M4 | N | NA | N | NA |
| E04 | 32 | M | R | N | PGY4 | Y | 6 | Y | 6 |
| N07 | 23 | F | R | Y | M1 | Y | 1 | Y | 1 |
| E05 | 31 | F | R | N | PGY5 | Y | 5 | Y (FLS) | 4 |
| N08 | 24 | M | R | Y | M1 | N | NA | N | NA |
| N09 | 24 | F | R | Y | M3 | N | NA | N | NA |
| E06 | 29 | F | R | N | PGY3 | Y | 3 | Y | 3 |
| E07 | 29 | M | R | N | PGY3 | Y | 3 | Y | 3 |
| E08 | 32 | F | R | N | NA | Y | 4 | Y | 4 |
| E09 | 33 | F | R | N | PGY6 | Y | 6 | Y | 4 |
| E10 | 30 | F | R | N | PGY5 | Y | 4 | Y | 4 |
| E11 | 42 | F | R | N | NA | Y | 13 | Y | 13 |
| E12 | 33 | M | R | N | NA | Y | 6 | Y | 6 |
| E13 | 49 | M | L | N | NA | Y | 20 | Y | 5 |
| E14 | 29 | F | R | N | PGY1 | Y | >1 | Y | >1 |
| E15 | 28 | M | R | N | PGY2 | Y | 2 | Y | 1 |
| N10 | 24 | F | R | Y | M1 | N | N | N | N |
| N11 | 23 | M | R | Y | M2 | N | N | N | N |
| N12 | 33 | F | R | N | N | N | N | N | N |
| N13 | 21 | F | R | N | N | N | N | N | N |
| N14 | 24 | F | R | N | N | N | N | N | N |
| N15 | 21 | M | R | N | N | N | N | N | N |

**S. Table 2:** Connectivity list

| S.N. | EEG-directed functional connectivity |
|---|---|
| 1 | LPFC-->RPFC |
| 2 | LPFC-->LM1 |
| 3 | LPFC-->RM1 |
| 4 | LPFC-->SMA |
| 5 | RPFC-->LPFC |
| 6 | RPFC-->LM1 |
| 7 | RPFC-->RM1 |
| 8 | RPFC-->SMA |
| 9 | LM1-->LPFC |
| 10 | LM1-->RPFC |
| 11 | LM1-->RM1 |
| 12 | LM1-->SMA |
| 13 | RM1-->LPFC |
| 14 | RM1-->RPFC |
| 15 | RM1-->LM1 |
| 16 | RM1-->SMA |
| 17 | SMA-->LPFC |
| 18 | SMA-->RPFC |
| 19 | SMA-->LM1 |
| 20 | SMA-->RM1 |

(a) 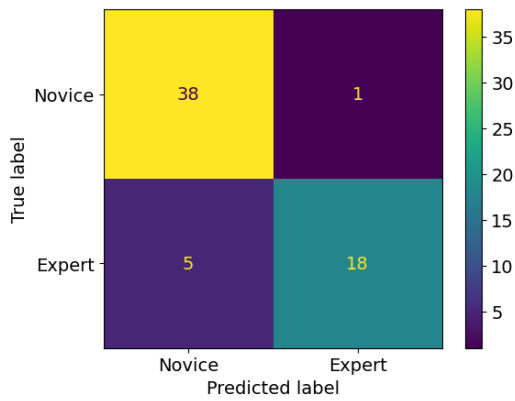     (b) 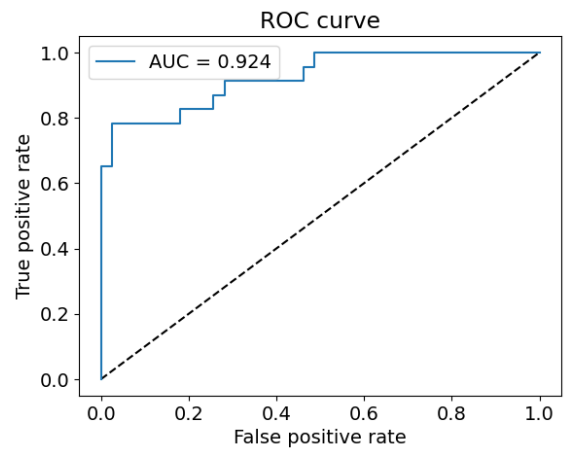

**S. Figure 3:** (a)Confusion matrix and (b)ROC curve for coarse subtask while picking up the suture.

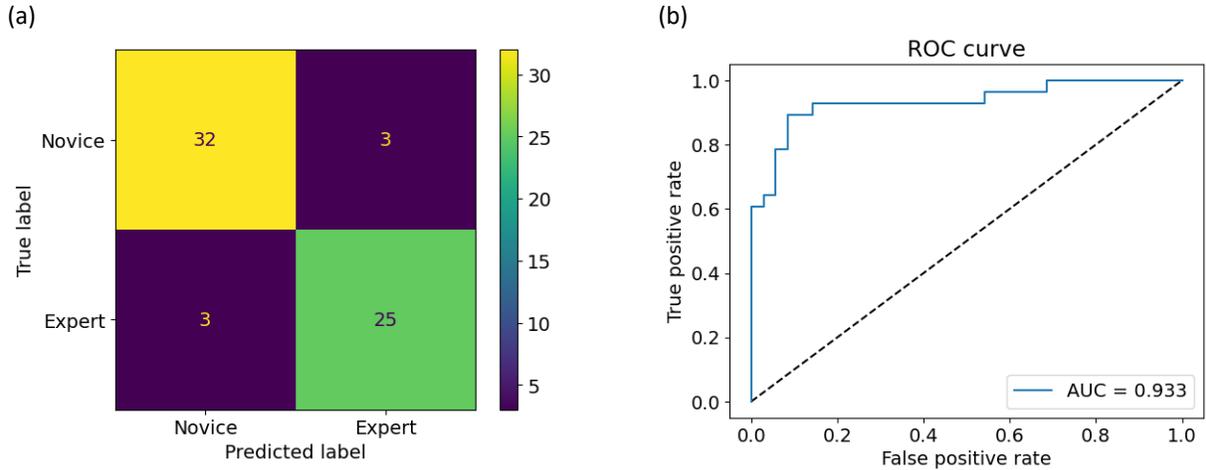

**S. Figure 4:** (a)Confusion matrix and (b)ROC curve for coarse subtask while inserting the needle through the Penrose drain.

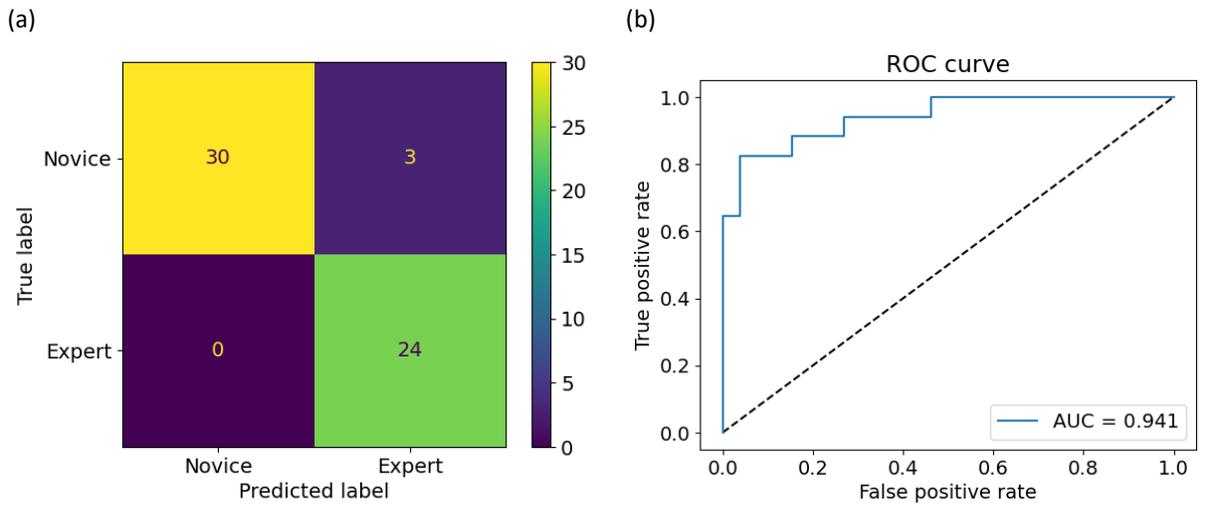

**S. Figure 5:** (a)Confusion matrix and (b)ROC curve for coarse subtask while tying three knots.

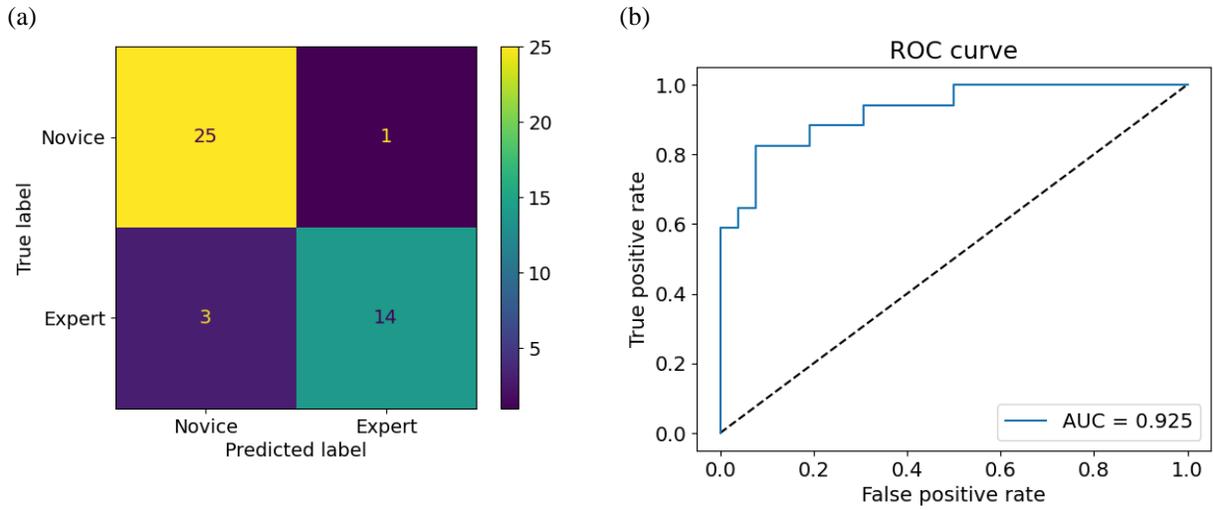

**S. Figure 6:** (a)Confusion matrix and (b)ROC curve for coarse subtask while cutting the suture.

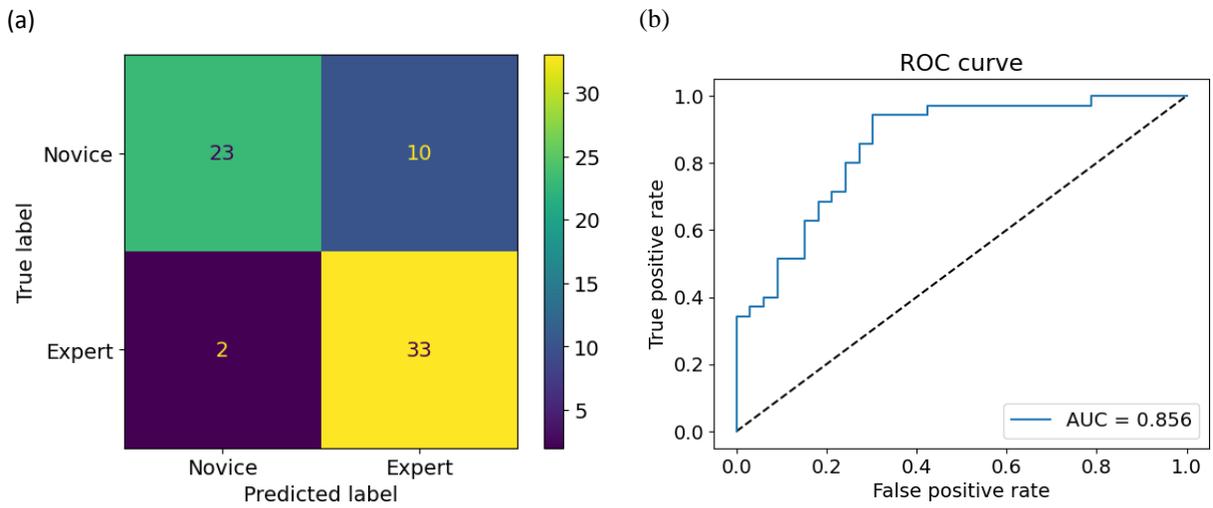

**S. Figure 7:** (a)Confusion matrix and (b)ROC curve for fine-level subtask picking up the suture.

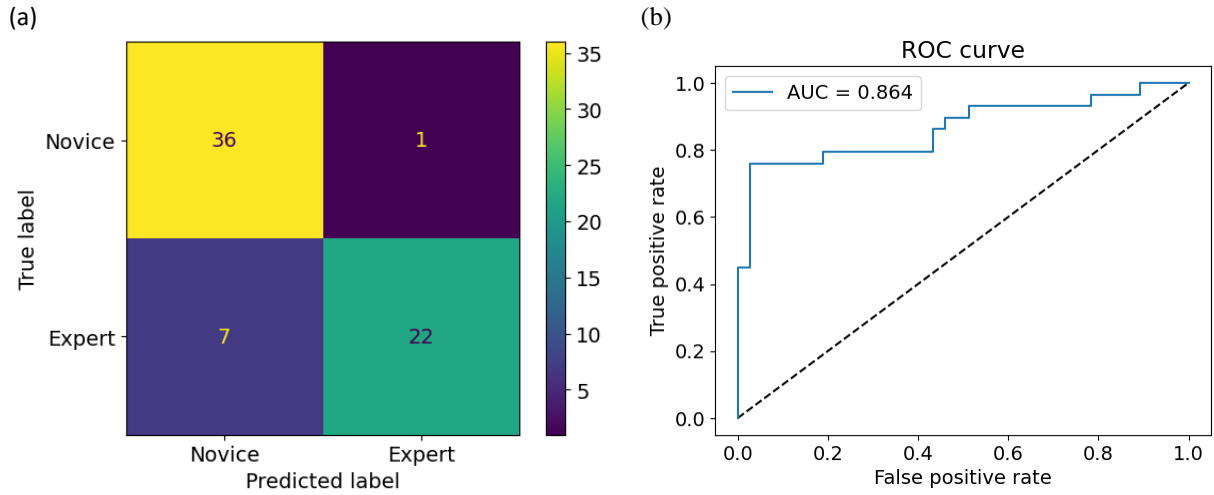

**S. Figure 8:** (a)Confusion matrix and (b)ROC curve for fine-level subtask introducing suture into the FLS trainer box.

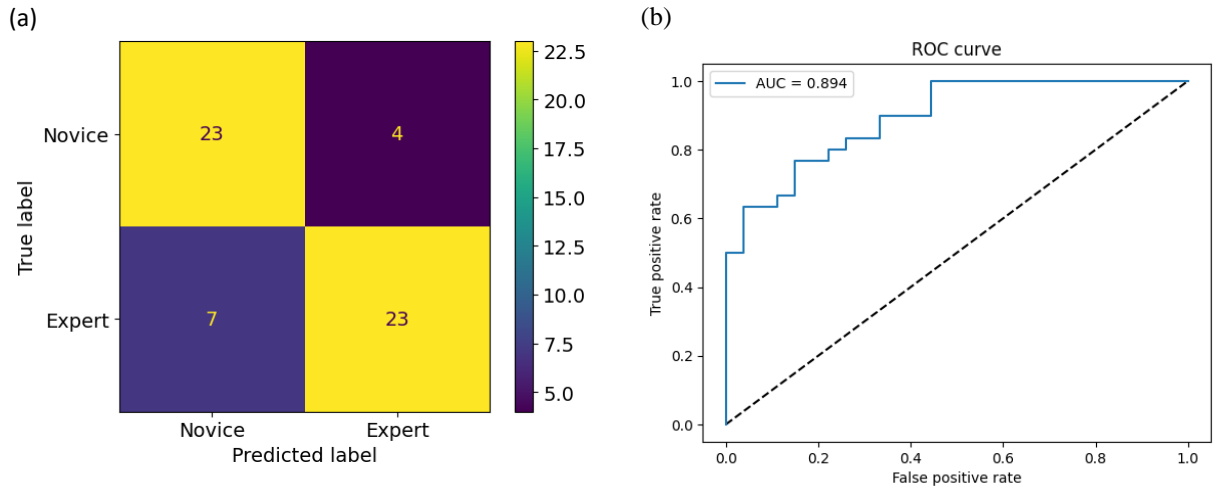

**S. Figure 9:** (a)Confusion matrix and (b)ROC curve for fine-level subtask orient the needle to the optimal direction for insertion in the Penrose drain.

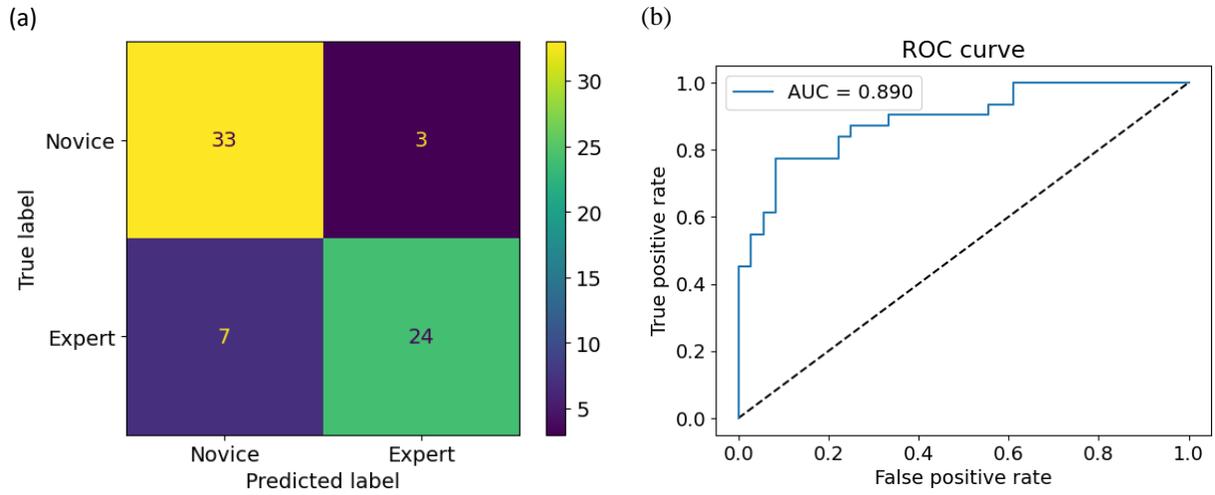

**S. Figure 10:** (a)Confusion matrix and (b)ROC curve for fine-level subtask insert the needle through two marks on the Penrose drain.

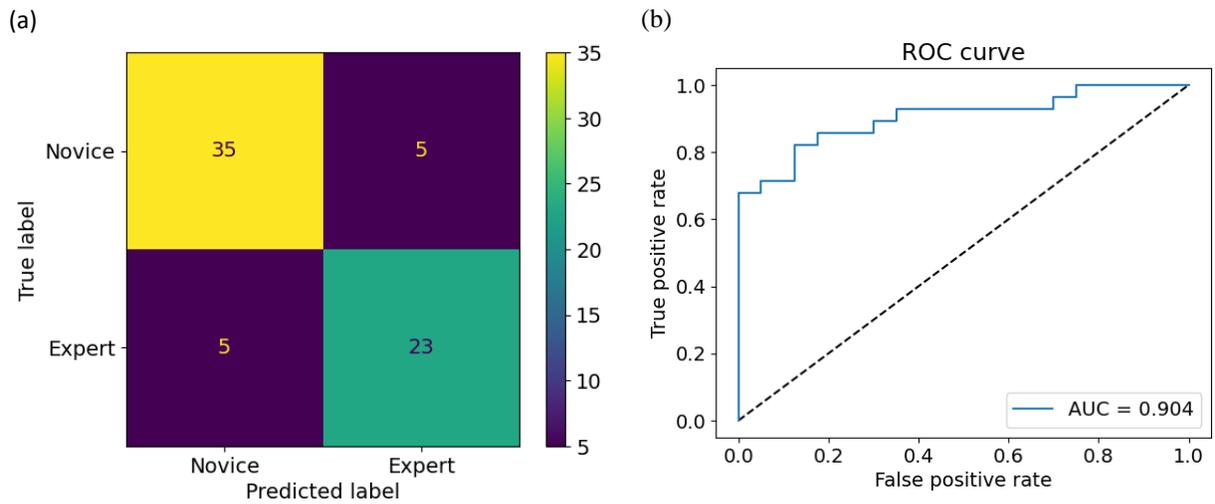

**S. Figure 11:** (a)Confusion matrix and (b)ROC curve for fine-level subtask pull the needle out of the Penrose drain.

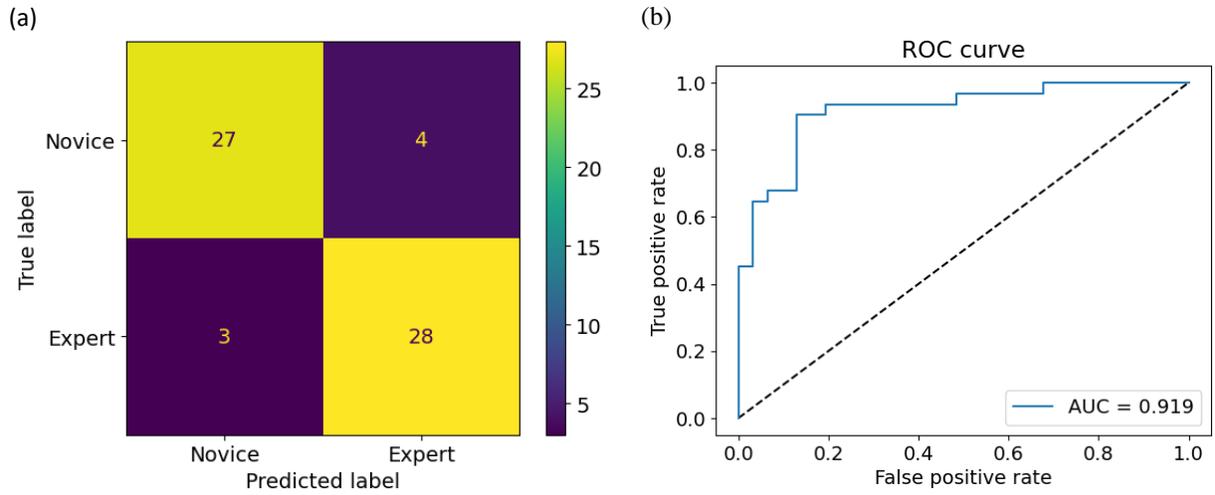

**S. Figure 12:** (a)Confusion matrix and (b)ROC curve for fine-level subtask make a double throw knot.

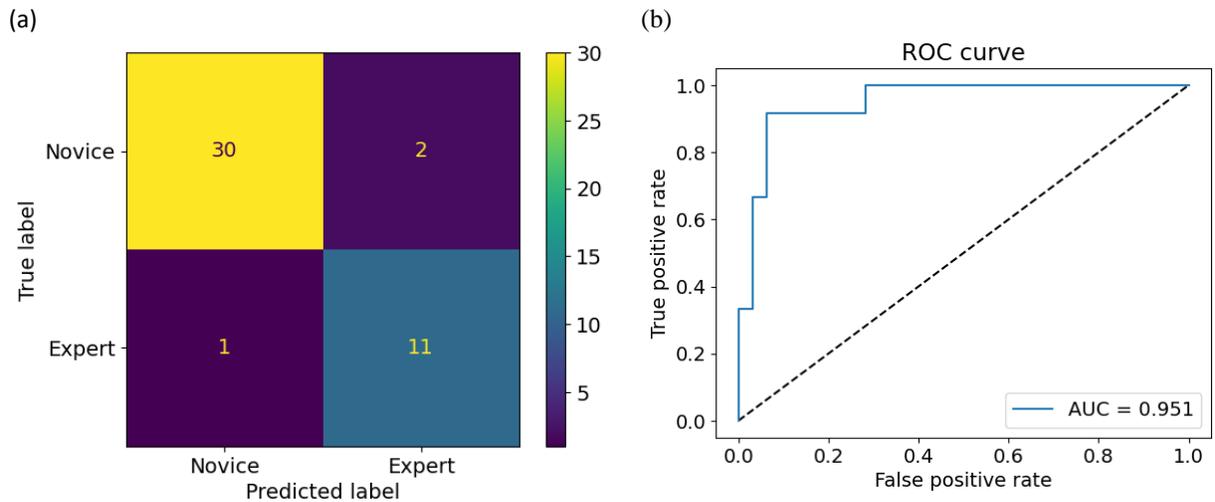

**S. Figure 13:** (a)Confusion matrix and (b)ROC curve for fine-level subtask switch the needle to the opposite hand.

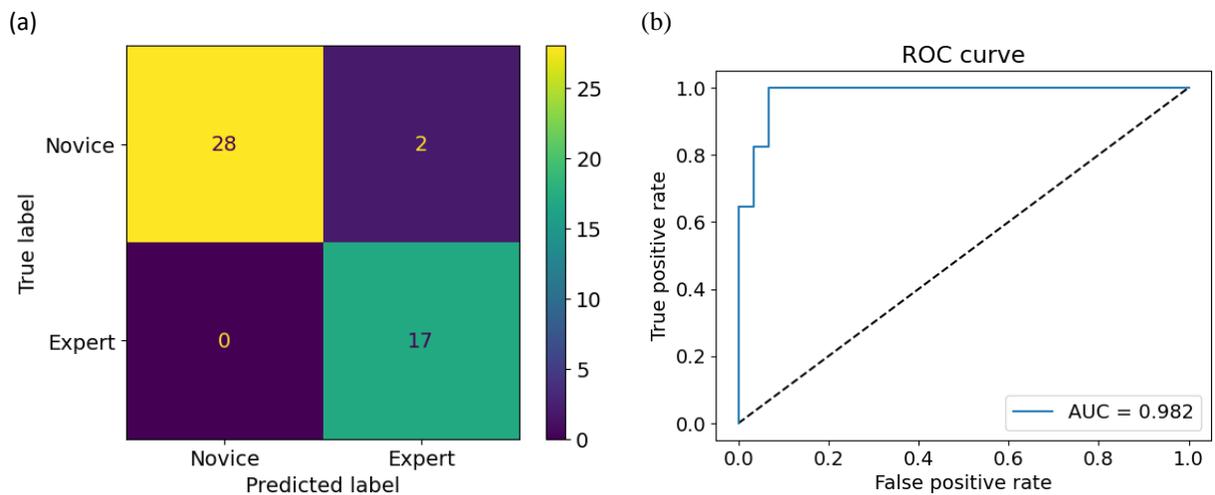

**S. Figure 14:** (a)Confusion matrix and (b)ROC curve for fine-level subtask make a single throw knot.

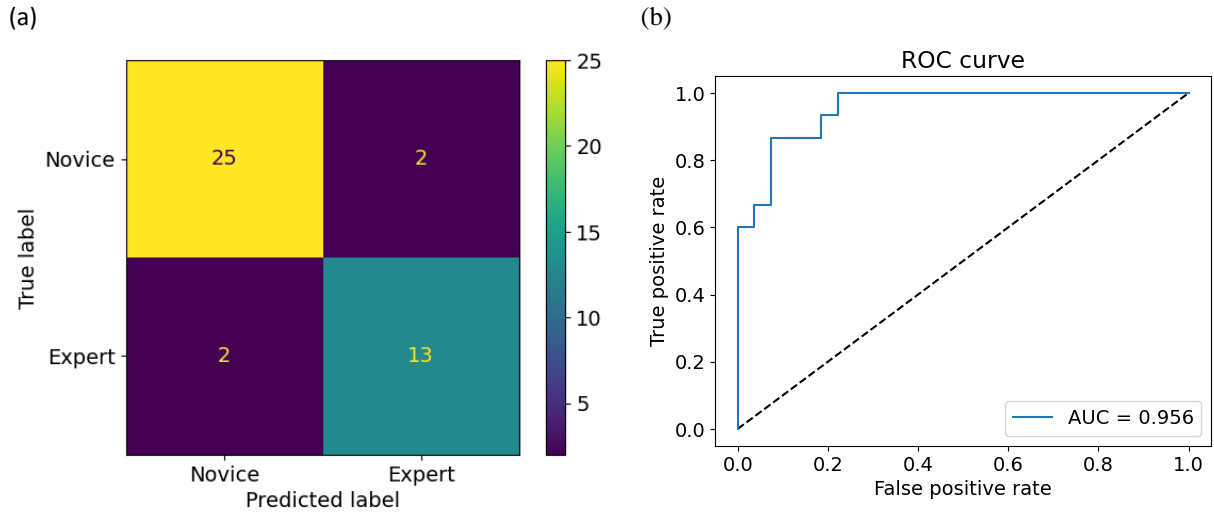

**S. Figure 15:** (a)Confusion matrix and (b)ROC curve for fine-level subtask while switching the needle to the opposite hand

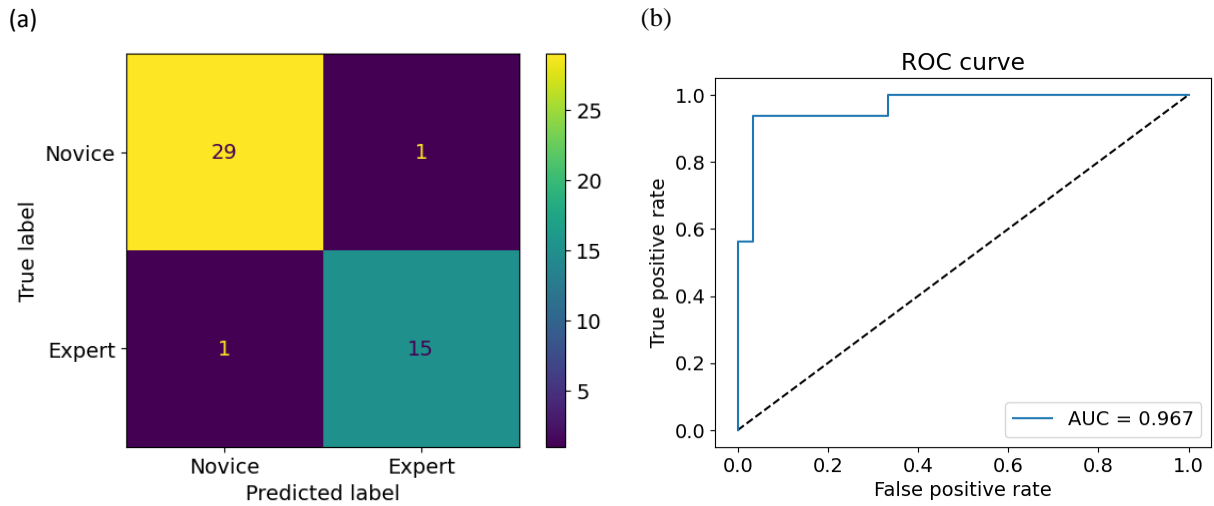

**S. Figure 16:** (a)Confusion matrix and (b)ROC curve for fine-level subtask while making first single throw knot.
(a)                                                         (b)

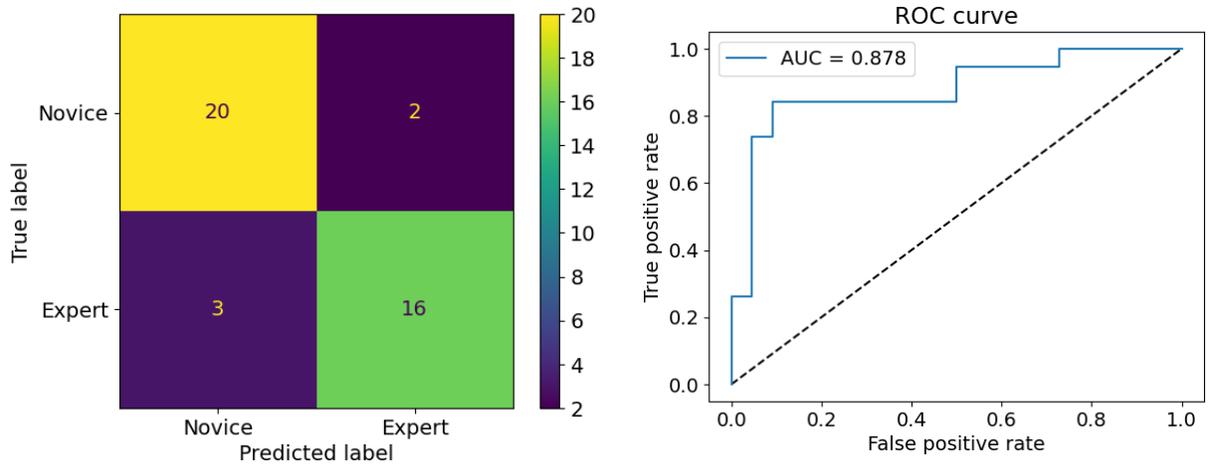

**S. Figure 17:** (a)Confusion matrix and (b)ROC curve for fine-level subtask while pulling out one of the needle drivers.

(a)                                                                                       (b)

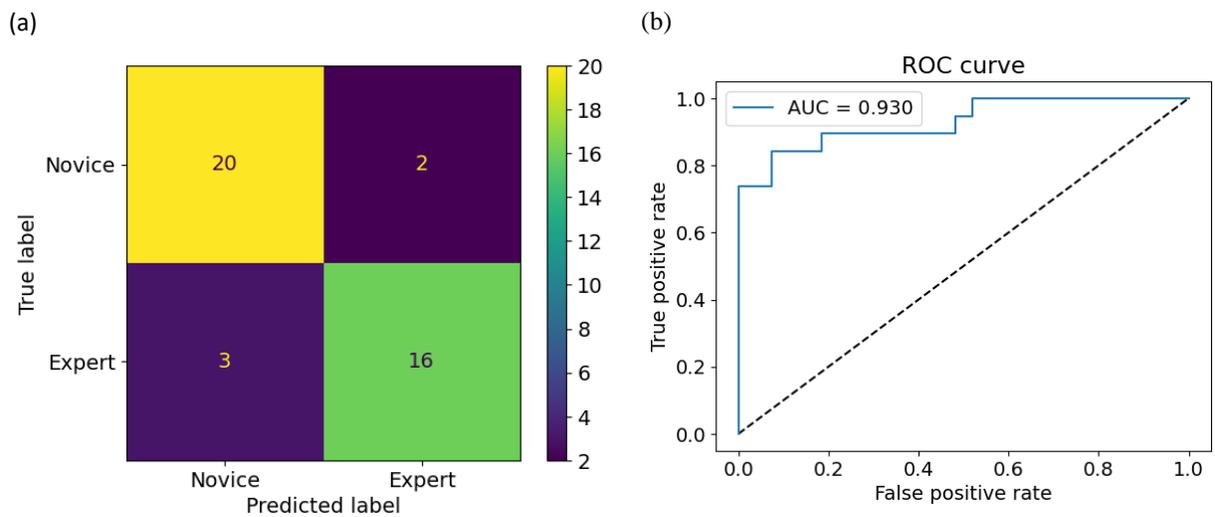

**S. Figure 18:** (a)Confusion matrix and (b)ROC curve for fine-level subtask introduce the scissors into the box.

(a)                                                                                       (b)

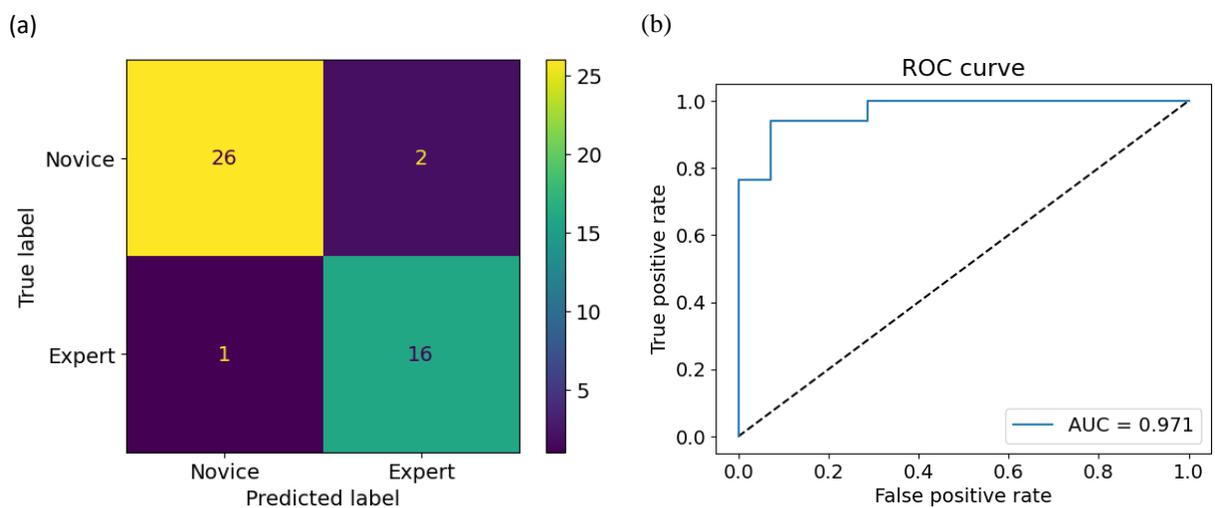

**S. Figure 19:** (a)Confusion matrix and (b)ROC curve for fine-level subtask while cutting both ends of the suture.

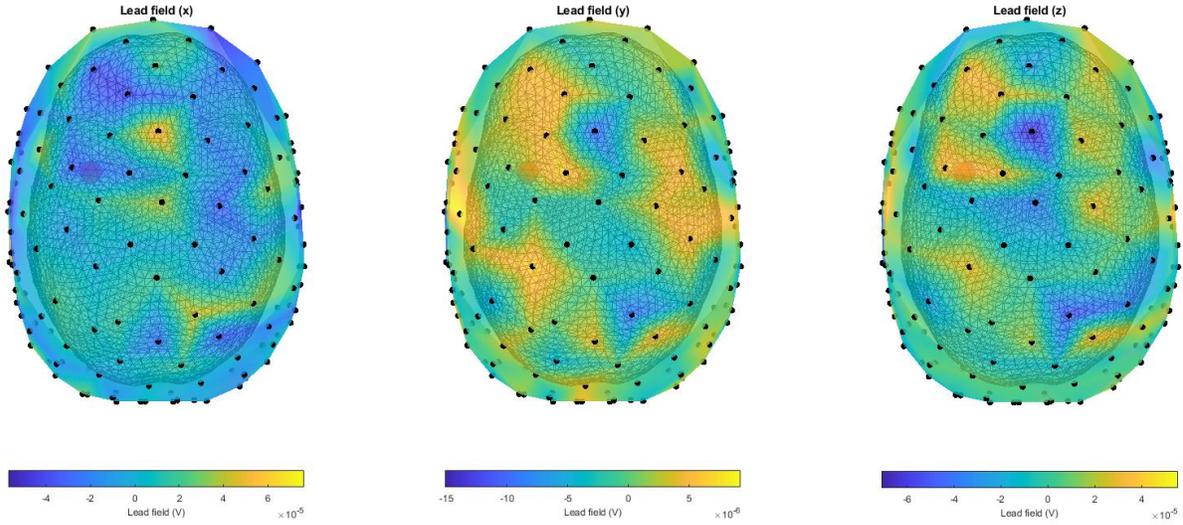

**S. Figure 20:** Lead field of EEG montage

## Non-linear Granger causality

Despite EEG brain signals being non-linear[146], the linear Granger causality (GC) framework is prominently applied for the computation of dFC in the brain due to its computational simplicity. However, linear GC requires stationarity of the signals and is inadequate for detecting non-linear causality[147]. In this study, we develop a novel method for computing non-linear Granger causality (nGC) by substituting the autoregressive (AR) component of linear GC with an attention-based long short-term memory (LSTM) encoder-decoder model. The details of our non-linear GC (nGC) method are described in the methods section. We evaluated our model on simulated linear (see equation S1) and non-linear (see equation S2) dynamical systems and compared its performance with the linear GC method. The results are summarized in S.Tables 3 and 4, respectively.

**Simulated linear system:**

The simulated linear dynamical system of signals ($x_1$, $x_2$, and $x_3$) is described as

$$x_1(t) = 0.5x_2(t-1) + 0.5x_3(t-1) + \epsilon_1(t-1)$$
$$x_2(t) = 0.5x_1(t-1) + 0.5x_2(t-1) + \epsilon_2(t-1)$$
$$x_3(t) = 0.99x_1(t-1) + \epsilon_3(t-1)$$
$$\epsilon_i \sim N(\mu = 0, \sigma = 0.01), i = 1,2,3$$

(S1)

In equation S1, 0.5, 0.99, etc. are the regression coefficients and $\epsilon_i$ is the additive white noise. The nGC and linear GC coefficients determined by the corresponding methods are shown in S.Table 3(a) and 3(b) respectively.

(a)

| Target | | Source | | |
|---|---|---|---|---|
| | | $x_1$ | $x_2$ | $x_3$ |
| | $x_1$ | 0.344 | 2.913 | 2.982 |
| | $x_2$ | 2.728 | 2.714 | 0.002 |
| | $x_3$ | 3.86 | 0.014 | 0.018 |

(b)

| Target | | Source | | |
|---|---|---|---|---|
| | | $x_1$ | $x_2$ | $x_3$ |
| | $x_1$ | NaN | 0.320 | 0.527 |
| | $x_2$ | 0.617 | NaN | 0.0001 |
| | $x_3$ | 1.468 | 0.00003 | NaN |

**S.Table 3:** Causality coefficients with (a) non-linear Granger causality (b) linear Granger causality

For the linear system, strong causal relationships are characterized by large regression coefficients, and they are assigned proportionally higher nGC coefficients by our model. For example, for source $x_1$ in S.Table 3(a), the targets $x_2$ and $x_3$ have significantly higher nGC coefficients, consistent with equation S1. The nGC coefficient for the target $x_1$ is comparatively lower since its associated regression coefficient is zero in equation S1 for the prediction of $x_1$ (t). Similarly, the nGC coefficients for the other source-target pairs are effectively captured by our model, as shown in S.Table 3(a). The linear GC method is also able to represent the relative causal relationship of the system as shown in S.Table 3(b). Typically, the linear GC coefficients are not computed for the same source ($x_i$) and target ($x_i$) which is indicated by NaN. It should be noted that the linear or non-linear Granger causality coefficients remain strictly non-negative, and a zero coefficient indicates non-causality[148].

**Simulated non-linear system:**

The simulated non-linear dynamical system of signals ($x_1$, $x_2$, and $x_3$) is described as

$$x_1(t) = 0.8x_1^2(t-3) + 0.2x_2^3(t-3) + 0.7x_3(t-3) + \epsilon_1(t-3)$$
$$x_2(t) = 3x_1^3(t-3) + 0.9x_2(t-3) + 0.5x_3^2(t-3) + \epsilon_2(t-3)$$
$$x_3(t) = 0.3x_2(t-3) + 0.3x_3^2(t-3) + \epsilon_3(t-3)$$
(S2)

$\epsilon_i \sim N(\mu=0, \sigma=0.01)$

The regression coefficients and additive white noise in equation S2 are similar to those defined for equation S1.

(a)

| | | Source | | |
|---|---|---|---|---|
| | | $x_1$ | $x_2$ | $x_3$ |
| Target | $x_1$ | 0.0165 | 0.0377 | 0.6158 |
| | $x_2$ | 0.0460 | 1.5835 | 0.154 |
| | $x_3$ | 0.0035 | 0.3677 | 0.0191 |

(b)

| | | Source | | |
|---|---|---|---|---|
| | | $x_1$ | $x_2$ | $x_3$ |
| Target | $x_1$ | NaN | 0.000026 | 0.443 |
| | $x_2$ | 0.0003 | NaN | 0.001 |
| | $x_3$ | 0.00003 | 0.281 | NaN |

**S.Table 4:** (a) non-linear Granger causality (b) linear Granger causality

For the non-linear system as well, our model assigned proportionally higher nGC coefficients for strong causal relationships that are characterized by large regression coefficients. For example, for source $x_1$ in S.Table 4(a), the nGC coefficients are $x_2$ which is proportional to regression coefficients in equation S2. Further, $x_1$ does not affect the prediction of $x_3$ (equation S2) which is accurately captured by our model with a low nGC coefficient (see S.Table 4(a), source $x_1$ target $x_3$). The causal relationships of other source-target pairs are also accurately elucidated by our model, as shown in S.Table 4(a). Although the linear GC method successfully identifies causality in linear systems, it couldn't evaluate the relative strength of relationships for the non-linear system. In particular, the linear GC coefficients for source-target pairs ($x_1$, $x_2$), ($x_2$, $x_1$), and ($x_3$, $x_2$) are not consistent with the regression coefficients in equation S2. These results highlight the robust characteristics of our non-linear GC method in capturing the strength of causality in linear and non-linear systems.

The performance of the LSTM full model for linear equation S1 is shown in S. Figure 21(a). The residuals of the prediction for all the variables were found to be normally distributed with a mean zero (see S. Figure 21(a)) and the autocorrelation was found to be high only at zero lag position (see S. Figure 21(b)) which delineates the high accuracy of the model prediction.

The effect of the removal of $x_1$, $x_2$, and $x_3$ in reduced-models can be seen in S. Figure 22(a-c). The reduced model-1 (after removal of past values $x_1$) struggled to accurately predict the future value of the variable $x_2$ and $x_3$ due to their strong dependency on past values of $x_1$. The removal of $x_1$ didn't affect its future value prediction (see Figure 22(a)). In other words, when the non-causal variables were removed, the prediction of the target variables was unaffected. Similarly, the causality of the other source-target pairs is also captured successfully by the LSTM model (S. Figure 22 (a-c)).

(a)

(a)

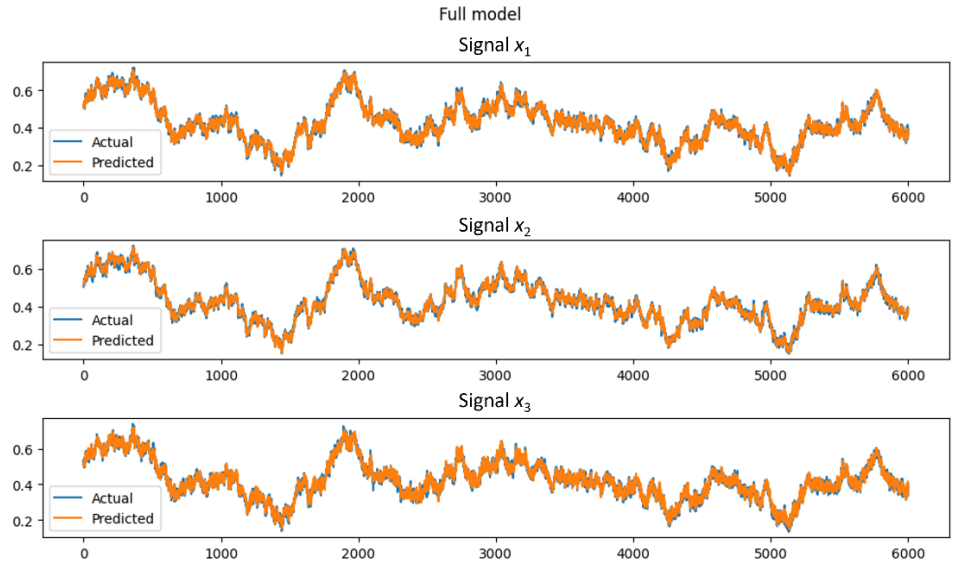

(b)

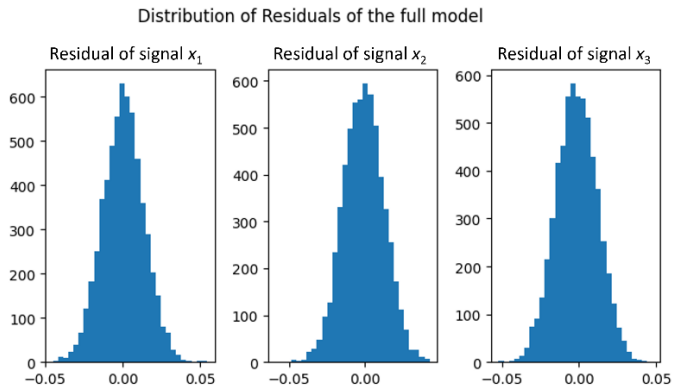

(c)

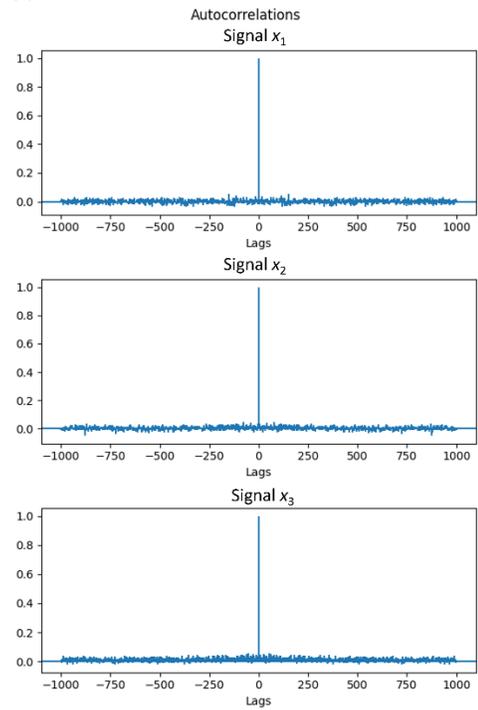

**S. Figure 21:** (a) The full-model was able to predict the future value of signals based on the past value by learning the causal relationship between them. (b) The residual of the prediction followed normal distribution indicating the high accuracy of the model. (c) The autocorrelations of the residuals are only correlated at zero lag which confirms the high modeling accuracy.

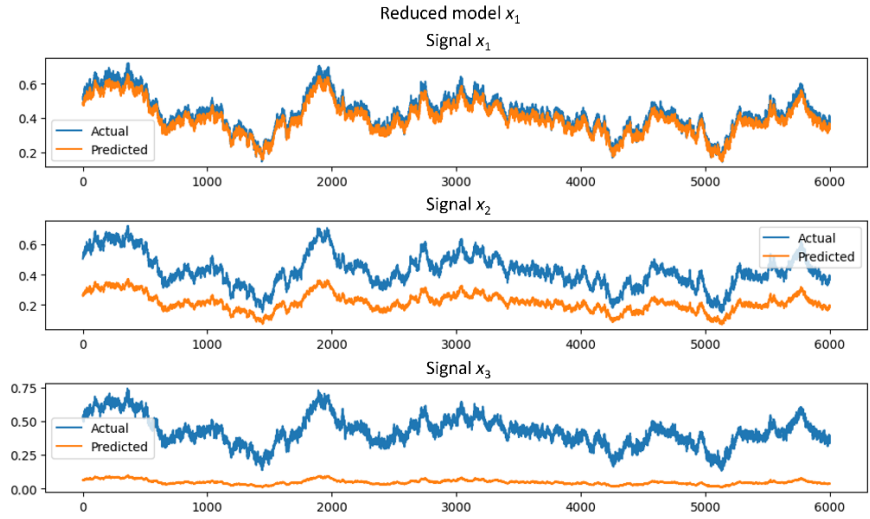

(b)

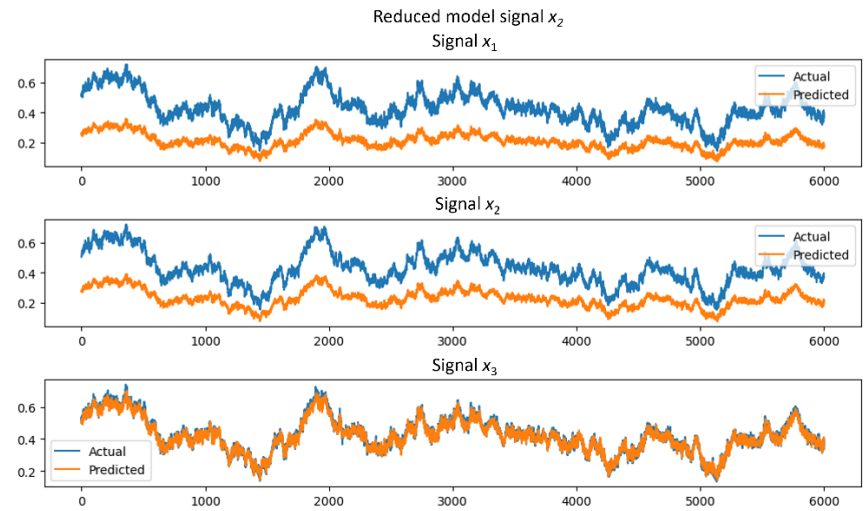

(c)

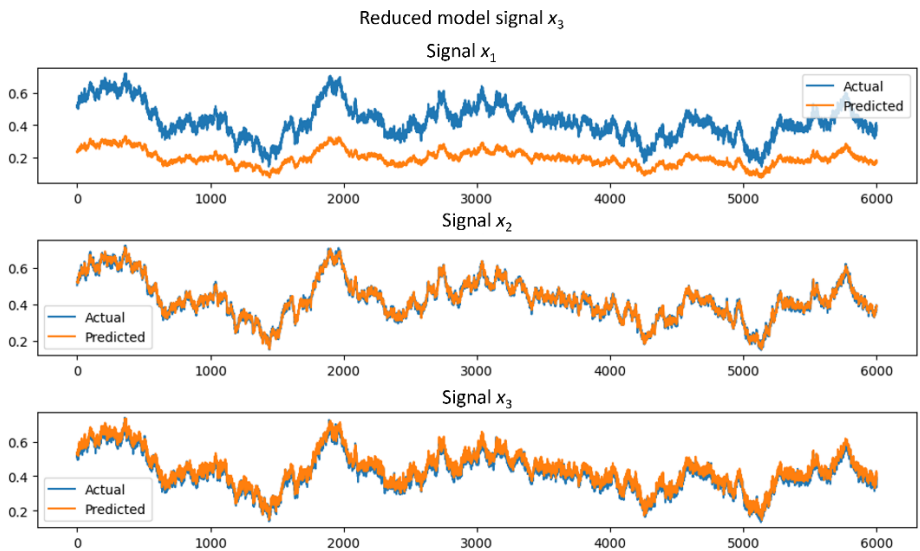

**S. Figure 22:** Prediction by the reduced model after (a) removing signal $x_1$ (b) removing signal $x_2$ (c) removing signal $x_3$.

**Non-Linear system:**

The non-linear Granger causality successfully highlighted the non-linear relationships among the signals (see S.Table 4). The causality analysis revealed that the model could detect complex dependencies.

The performance of the LSTM full model is shown in S. Figure 23(a). As can be seen in this figure, the full-model was able to predict the future timeseries values based on the past values by learning the internal dynamics of the relationship between them. The residuals of the prediction for all the variables were found to be normally distributed with a mean about zero (see S. Figure 23(b)) and the autocorrelation was found to be high only at zero lag (see S. Figure 23(c)) which delineates the high accuracy of the model prediction.

The effect of the removal of $x_1$, $x_2$, and $x_3$ in reduced-models can be seen in S.Figure 24(a-c). The reduced model-2 (after removal of past values $x_2$) struggled to accurately predict the future value of the variable $x_1$, $x_2$ and $x_3$ due to their strong dependency on past values of $x_2$. The strong dependency is given a high nGC coefficient by the LSTM model (see S.Table 4(a), source x₂ target $x_1$, $x_2$ and $x_3$) which are proportional to their inter-dependent coefficients in equation S2. Similarly, the causality of the other source-target pairs is also captured successfully by the LSTM model which can be seen in S.Table (2) and the effect of the removal of the variable can be seen in S.Figure 24 (a-c).

(a)

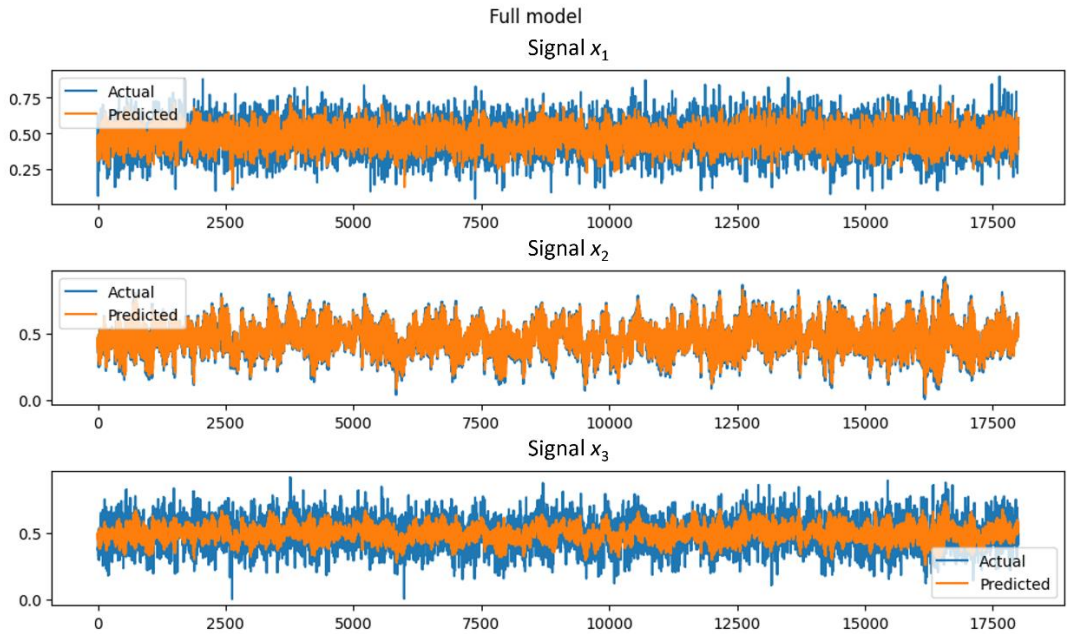

(b)  (c)

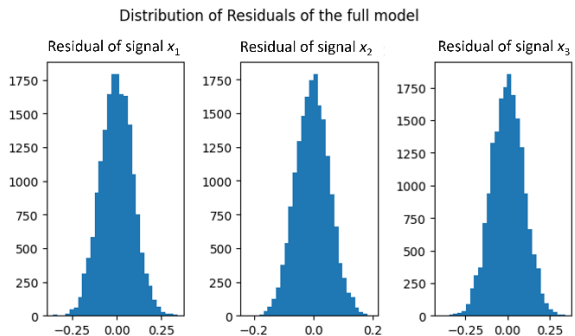

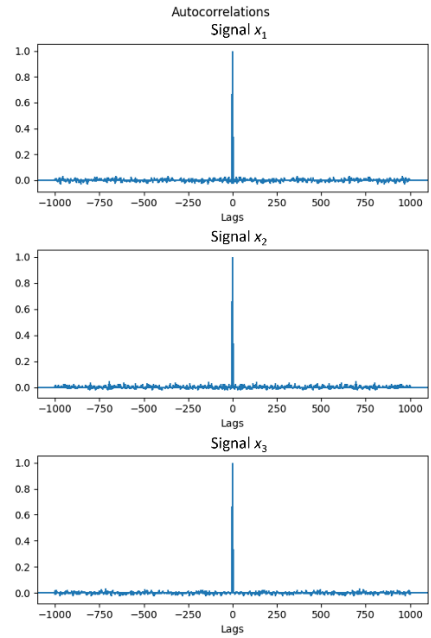

**S. Figure 23:** (a) The full-model was able to predict the future value of signals based on the past value by learning the causal relationship between them. (b)The residuals of the prediction followed normal distribution indicating the high accuracy of modeling the system. (c) The autocorrelations of the residuals are only correlated at zero lag which confirms the high accuracy of modeling.

(a)

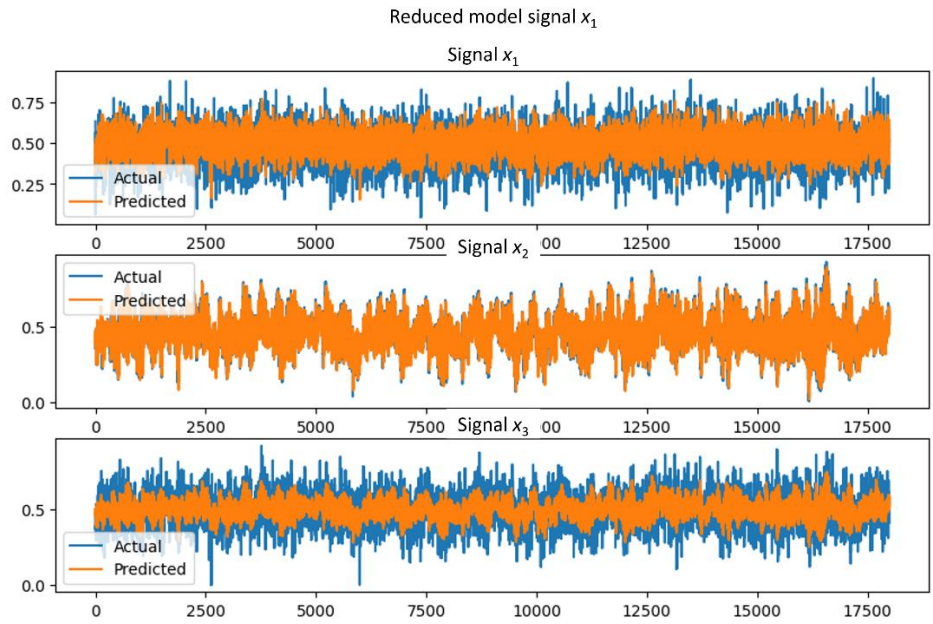

(b)

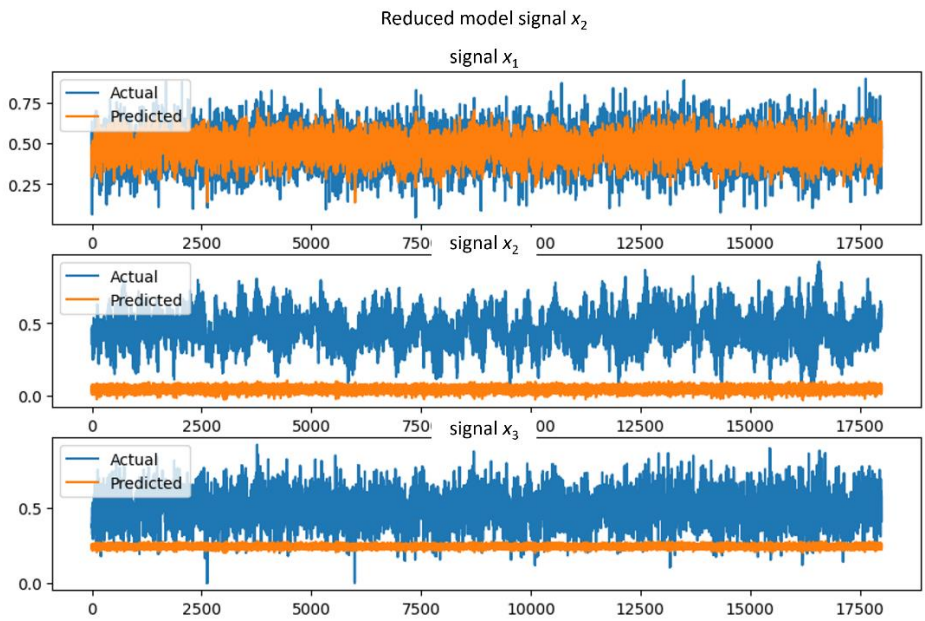

(c)

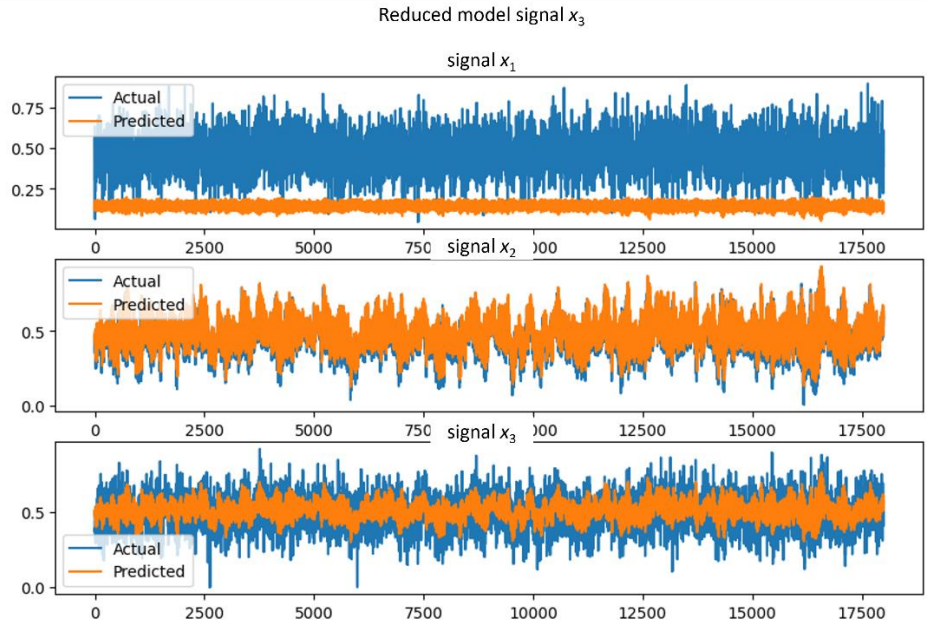

**S. Figure 24:** Prediction by the reduced model after (a) removing signal $x_1$ (b) removing signal $x_2$ (c) removing signal $x_3$.

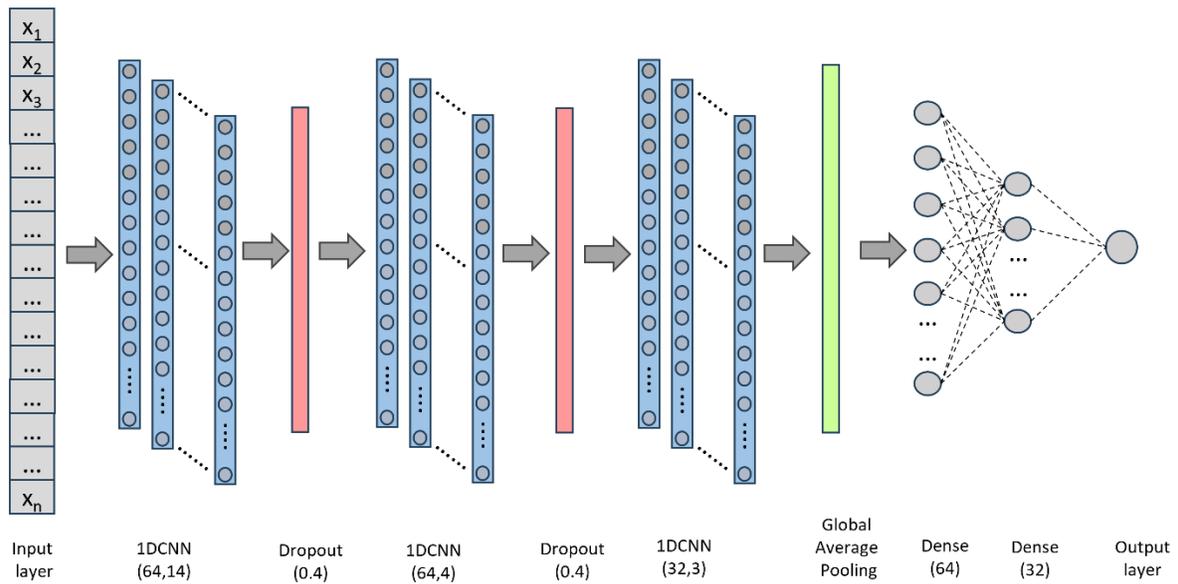

**S. Figure 25:** Architecture of the 1DCNN model